\pdfoutput=1
\documentclass[twocolumn,twocolappendix]{aastex631}
\usepackage{here}
\usepackage{url}
\usepackage{color,ulem}

\defcitealias{2019ApJ...876...27Y}{YL19}

\shorttitle{Non-thermal emission from SNRs in a diversified CSM}
\shortauthors{Kobashi et al.}
\graphicspath{{./}{figures/}}

\begin{document}

\title{Long-term evolution of non-thermal emission from Type Ia and core-collapse supernova remnants in a diversified circumstellar medium}

\author{Ryosuke Kobashi}
\affiliation{Department of Astronomy, Kyoto University, Kitashirakawa, Oiwake-cho, Sakyo-ku, Kyoto 606-8502, Japan}

\author[0000-0002-0802-6390]{Haruo Yasuda}
\affiliation{Department of Astronomy, Kyoto University, Kitashirakawa, Oiwake-cho, Sakyo-ku, Kyoto 606-8502, Japan}

\author[0000-0002-2899-4241]{Shiu-Hang Lee}
\affiliation{Department of Astronomy, Kyoto University, Kitashirakawa, Oiwake-cho, Sakyo-ku, Kyoto 606-8502, Japan}
\affiliation{Kavli Institute for the Physics and Mathematics of the Universe (WPI), The University of Tokyo, Kashiwa 277-8583, Japan}

\correspondingauthor{Ryosuke Kobashi}
\email{kobashi@kusastro.kyoto-u.ac.jp}



\begin{abstract}
The contribution of galactic supernova remnants (SNRs) to the origin of cosmic rays (CRs) is an important open question in modern astrophysics. Broadband non-thermal emission is a useful proxy for probing the energy budget and production history of CRs in SNRs. We conduct hydrodynamic simulations to model the long-term SNR evolution from explosion all the way to the radiative phase (or $3\times10^5$ yrs at maximum), and compute the time evolution of the broadband non-thermal spectrum to explore its potential applications on constraining the surrounding environments as well as the natures and mass-loss histories of the SNR progenitors. A parametric survey is performed on the ambient environments separated into two main groups, namely a homogeneous medium with a uniform gas density and one with the presence of a circumstellar structure created by the stellar wind of a massive red-supergiant (RSG) progenitor star.  
Our results reveal a highly diverse evolution history of the non-thermal emission closely correlated to the environmental characteristics of a SNR. Up to the radiative phase, the roles of CR re-acceleration and ion-neutral wave damping on the spectral evolution are investigated. Finally, we make an assessment of the future prospect of SNR observations by the next-generation hard X-ray space observatory \textit{FORCE} and predict what we can learn from their comparison with our evolution models.      
%
\end{abstract}




\section{Introduction} \label{sec:intro}

Supernova remnants (SNRs) are believed to be an important source of cosmic rays (CRs) in our and other galaxies  \citep[e.g.,][]{1987PhR...154....1B,2005ApJ...632..920E}. To quantify the contribution of SNRs to the production of Galactic CRs, it is necessary to follow the production history of CRs in a SNR throughout its lifetime.
Observationally, non-thermal emission across a wide energy range covering the radio, X-rays and gamma-rays is a powerful tool for the inference of CR production in a SNR. The diverse interstellar medium (ISM) and circumstellar medium (CSM) surrounding the SNRs are known to be one of the most important determining factors for the CR acceleration history and hence the resulted time evolution of the non-thermal emission \citep[e.g.,][]{2021ApJ...919L..16Y,Yasuda_2022}. Comparisons of hydrodynamic models with observational data have been performed for individual SNRs to estimate their CR energy budgets \citep[e.g.,][]{2013ApJ...767...20L,Slane2014}. However, systematic parametric surveys taking into account the rich diversity of the ambient environments and progenitors of various types of SNRs are still lacking.  

The CR acceleration efficiency and thus the total amount of CR produced in a SNR strongly depends on the ambient environment, age as well as the progenitor system. Therefore, it is important to quantify the effects of environmental parameters such as the ambient gas density and magnetic field profiles using a self-consistent numerical setup.
Indeed, according to \citet[][]{2019ApJ...876...27Y} (hereafter \citetalias{2019ApJ...876...27Y}) who performed such a task up to a SNR age of 5,000 yrs, a rich variety of non-thermal emission evolution has been found under different parameters for the surrounding environments. Another aspect to be explored is the ability of non-thermal emission observations on constraining the CSM structure and hence the pre-SN mass-loss activities of SN progenitors. 

In this study, we extend the work by \citetalias{2019ApJ...876...27Y} to follow the evolution from explosion all the way to an age of a few $10^4$ to $3\times10^5$ yrs, as well as the implementation of more realistic CSM environments for a red supergiant (RSG) star by performing a hydrodynamic simulation for the pre-SN wind-ISM interaction. We simulate the shock hydrodynamics and CR acceleration simultaneously until the forward shocks have weakened enough to stop accelerating CRs efficiently up to an age of $\sim$ a few or tens of $10^4$ yrs depending on the ejecta/CSM model. The resulted grid of SNR/CSM models will provide a broader vision on the long-term evolution of non-thermal emissions from SNRs interacting with different kinds of environments. 

One of the novel aspects introduced in this paper is the self-consistent inclusion of the radiative phase of a SNR inside our CR-hydrodynamic simulation framework (a similar approach has been adopted in several previous studies, e.g., \citet{2015ApJ...806...71L,2020AA...634A..59B,2021AA...654A.139B}). 
Radio and gamma-ray bright middle-aged SNRs such as W44 and IC 443 \citep[e.g.,][]{2013Sci...339..807A} are usually found to possess radiative shocks. Their bright radio synchrotron emission and GeV gamma-rays from $\pi^0$-decay are suggested to originate from a rapid compression of gas, CRs and magnetic field in the cold dense shell formed behind the radiative shocks \citep[e.g.,][]{2015ApJ...806...71L}.
In the radiative phase, it has been suggested that re-acceleration of pre-existing CRs \citep[e.g.,][]{2010ApJ...723L.122U} and the effects of ion-neutral damping of CR-trapping magnetic waves \citep[e.g.,][]{2011NatCo...2..194M,2013SSRv..178..599B} are important to account for the observed spectral properties in evolved SNRs. 
Some numerical studies have taken into account the effects of re-acceleration and wave damping at shocks propagating in dense environments such as molecular clouds and becoming radiative so that the CR acceleration efficiency is no longer high \citep[e.g.,][]{2015ApJ...806...71L,galaxies7020049}. An alternative interpretation using a CR escape model \citep[e.g.,][]{2009MNRAS.396.1629G,2010AA...513A..17O} has also been proposed, which can explain such characteristic spectra from evolved SNRs by a rapid decrease of the maximum proton energy with age 
\citep{2019MNRAS.487.3199C,2020AA...634A..59B}. \citet{2015ApJ...806...71L} calculated the hydrodynamics of a fast cloud shock driven into a dense cloud by a SNR and the accompanying non-thermal emission. Likewise, \citet{galaxies7020049} performed a similar calculation but used an analytic approach for the hydrodynamics. 
Both works did not survey over different SN progenitors or CSM models. They also did not consider the SNR evolution and particle acceleration in the free-expansion and Sedov phase before the shock-cloud interaction begins, which can have a non-negligible contribution to the non-thermal emission even at old ages. 
A few other previous theoretical studies have also investigated the evolution of CR energetics as a function of SNR age \citep[e.g.,][]{LEN2012}, but these works primarily focused on the younger remnants without any discussion on the later evolution stages. Our study addresses these points using a coherent hydrodynamic simulation to connect the youngs to the olds.  
In addition, from the view point of better understanding the connection between the non-thermal emission properties of young and evolved SNRs,
it is important to understand the role of re-acceleration of pre-existing CRs as a function of age under different ambient environment settings. In this study, using our grid of SNR/CSM models from explosion up to the radiative phase, we quantify the importance of CR re-acceleration in terms of the total CR energy budget throughout the lifetime of a SNR. 


In the last part of the paper, we assess the future prospect of \textit{FORCE} (\textit{Focusing On Relativistic universe and Cosmic Evolution}, \url{https://www.cc.miyazaki-u.ac.jp/force/wp-content/uploads/force_proposal.pdf}), a next-generation hard X-ray imaging observatory, on constraining particle acceleration parameters using our grid of SNR/CSM models. When it comes to X-rays, observations of various SNRs have been done in the soft X-ray bands using instruments on board satellites such as \textit{Chandra}, \textit{Suzaku} and \textit{XMM-Newton}, for which there is often much contamination from the thermal emission when one tries to separate out the non-thermal component. For the study of CR (electron) acceleration, hard X-ray data at $> 10$ keV with good statistics is highly desirable. Such observations have been performed for a few examples using the \textit{NuSTAR} observatory with an arcmin scale spatial resolution which is close to the angular size of many young SNRs. The power-law index of CR electrons can be constrained from the high-energy edge of the synchrotron tail for some SNRs, e.g., RX J1713.7-3946 \citep[][]{2019ApJ...877...96T} and Tycho's SNR \citep[][]{2015ApJ...814..132L}, allowing one to constrain the acceleration efficiency of electrons and magnetic field strengths, as well as the non-thermal bremsstrahlung emission from some SNRs interacting with dense clouds, e.g., W49B \citep[][]{2018ApJ...866L..26T} and IC443 \citep[][]{2018ApJ...859..141Z}, providing information on the sub-relativistic accelerated particles and hence the poorly understood electron injection process. 
Here we expect future observations using the \textit{FORCE} satellite which is planned to launch in the later half of 2020's and will observe SNRs with a high sensitivity in the 10--40 keV band. With an angular resolution $<15''$ which is $\gtrsim4$ times better than \textit{NuSTAR}, \textit{FORCE} will enable us to realize spatially-resolved spectroscopic observations of SNRs in the crucial hard X-ray window.  

This paper is structured as follows. In Section~\ref{sec:methods} we first explain our numerical methods which enable us to calculate SNR evolution until a few or tens of $10^4$ yrs, and then introduce our models for the surrounding environments in this paper, i.e., models with a uniform ambient medium and those with a CSM created by the pre-SN stellar wind. In Section~\ref{subsec:res-uni} and Section~\ref{subsec:res-csm}, we present our results from both classes of models sequentially and discuss their various implications. Section~\ref{subsec:res-rad} is dedicated to the analyses of a few physical effects especially relevant to the non-thermal emission in the radiative phase, followed by a brief discussion on the future prospect of \textit{FORCE} in hard X-ray studies of young and old SNRs in Section~\ref{subsec:res-force}. Section~\ref{sec:summary} provides a summary of our results and concluding remarks.   


\section{Methods} \label{sec:methods}

\subsection{Included physics}

We use the \textit{CR-Hydro} code developed by \citetalias{2019ApJ...876...27Y} with adaptations to fit the purposes of this work. The \textit{CR-Hydro} code performs 1-D spherically symmetric hydro simulations on a Lagrangian grid \textit{VH-1} \citep[e.g.,][]{2001ApJ...560..244B} coupled with a semi-analytic non-linear diffusive shock acceleration (NLDSA) calculation \citep[e.g.,][]{2004JKAS...37..483B, 2010APh....33..307C, 2010MNRAS.407.1773C} similar to the framework introduced in e.g., \citet{LEN2012}. To account for the feedback of the accelerated particles and magnetic fields on the hydrodynamics, the code uses an effective ratio of specific heats $\gamma_\mathrm{eff}$ which is updated in real-time at each Lagrangian cell as follows \citep[][]{2001ApJ...560..244B}, 

\begin{equation}
\frac{\gamma_\mathrm{eff}}{\gamma_\mathrm{eff}-1}P_\mathrm{tot}=\frac{\gamma_\mathrm{g}}{\gamma_\mathrm{g}-1}P_\mathrm{g}+\frac{\gamma_\mathrm{CR}}{\gamma_\mathrm{CR}-1}P_\mathrm{CR}+\frac{\gamma_\mathrm{B}}{\gamma_\mathrm{B}-1}P_\mathrm{B},
\end{equation}
where $\gamma_\mathrm{g}=5/3, \gamma_\mathrm{CR}=4/3, \gamma_\mathrm{B}=2$ is the ratio of specific heats for ideal gas, CR, magnetic field respectively, $P_\mathrm{tot}=P_\mathrm{g}+P_\mathrm{CR}+P_\mathrm{B}$ is the total pressure and $P_\mathrm{g}, P_\mathrm{CR}, P_\mathrm{B}$ are gas, CR, and magnetic pressures, respectively. 

For radiative cooling, we adopt the non-equilibrium (NEQ) cooling function from \citet{1993ApJS...88..253S} coupled to the exact time integration method of \citet{2009ApJS..181..391T}. 
In accordance with \citet{1998ApJ...500..342B}, we introduce the timescale $t_\mathrm{tr}$ for the transition to the radiative phase \footnote{By ``radiative phase'' we refer to the age when the post-shock radiative cooling effect becomes important on the shock dynamics. We are duly noted that this is different from the conventional definition of the radiative phase which is when $R_\mathrm{sk}\sim t^{2/7}$ holds after the shock oscillation has subsided, as in \citet{2021MNRAS.505..755P}.} which will be used as a basic time unit for our results throughout the paper, i.e., 
\begin{equation}
    t_\mathrm{tr}\approx 2.9\times10^4\ (E_\mathrm{SN}/1.0\times10^{51}\ \mathrm{erg})^{4/17}\ n_0^{-9/17}\ \mathrm{yr}
\end{equation}
where $E_\mathrm{SN}$ is the SN explosion energy and $n_0$ is the number density of the ambient gas in $\mathrm{cm}^{-3}$. 

To obtain the phase-space distribution function of the accelerated protons $f(x,p)$, we solve the diffusion-convection equation in the shock rest frame \citep[e.g.,][]{2004JKAS...37..483B,2010APh....33..307C,2010MNRAS.407.1773C,LEN2012}. From the formulation of the solution $f(x,p)$ \citep[][Eq.13]{LEN2012}, we can decompose it into two components depending on the type of seed particles being accelerated from: 
\begin{eqnarray}
\label{eq:solution}
    f(x&,&p)\propto\biggl[{\eta n_0\over 4\pi p_\mathrm{inj}^3}\exp\biggl(-\int_{p_\mathrm{inj}}^p{dp'\over p'}\frac{3S_\mathrm{tot}U(p')}{S_\mathrm{tot}U(p')-1}\biggr)\biggr]\\
    &+&\int_{p_\mathrm{inj}}^p{dp''\over p''}f_\mathrm{pre,p}(p'')\biggl[\exp\biggl(-\int_{p''}^p{dp'\over p'}\frac{3S_\mathrm{tot}U(p')}{S_\mathrm{tot}U(p')-1}\biggr)\biggr],\nonumber
\end{eqnarray}
where $S_\mathrm{tot}$ is the effective total compression ratio of the shock, $U(p)$ is the dimensionless gas flow velocity, $f_\mathrm{pre,p}$ is the distribution function of any pre-existing CR protons. Using the parameterization for the injection efficiency in the language of the so-called ``thermal leakage'' model, the fraction of downstream thermal particles being injected into the DSA process is $\eta=\{4/(3\sqrt{\pi})\}(S_\mathrm{sub}-1)\chi_\mathrm{inj}^3e^{-\chi_\mathrm{inj}^2}$ with $S_\mathrm{tot}=(u_0-v_{A,0})/(u_2+v_{A,2})$, $S_\mathrm{sub}=(u_1-v_{A,1})/(u_2+v_{A,2})$ based on the so-called ``Alfv\'enic drift" model (see Section~\ref{subsec:MFA} for a discussion on a few caveats along this line), where $u_i$ is the gas velocity and $v_{\mathrm{A},i}$ is the Alfv\'en speed, for which the subscript $i$ indicates values at far upstream (0), immediately upstream (1), immediately downstream (2) of shock respectively. The dimensionless quantity $\chi_\mathrm{inj} = 3.8$ is chosen to reflect the typical values inferred from emission modeling of a few young SNRs \citep{LEN2012,2013ApJ...767...20L,Slane2014} \footnote{We note that the actual DSA injection mechanism at the shock is not necessarily a ``thermal leakage'' process, but we are using the parameterization scheme for numerical convenience.}. To obtain the electron distribution function we assume an electron-to-proton number ratio at relativistic energies $K_\mathrm{ep} = 10^{-2}$ (c.f. model B in \citetalias{2019ApJ...876...27Y}). We calculate the maximum energy of the accelerated particles as the minimum of the age-limited, loss-limited (mainly for electrons) and escape-limited maximum energies at each time epoch. The same approach has been adopted in e.g.,  \citet{LEN2012,Slane2014,2019ApJ...876...27Y,2021ApJ...919L..16Y,Yasuda_2022}.

It has been suggested that re-acceleration of pre-existing CRs plays a pivotal role in the production of non-thermal emission in older SNRs  \citep{2010ApJ...723L.122U,2015ApJ...806...71L}. We will further elaborate in Section~\ref{subsec:res-reacc} on the mechanism in detail. 
Following \citet{2010ApJ...723L.122U, 2015ApJ...806...71L}, we assume that such pre-existing CRs have phase space distributions $f_\mathrm{pre,p/e}$ of the Galactic CR protons and electrons+positrons,
\begin{eqnarray}\label{eq:pre-reacc}
4\pi p^2f_\mathrm{pre,p}&=&4\pi J_p\beta^{1.5}p_0^{-2.76}\\ \nonumber
4\pi p^2f_\mathrm{pre,e}&=&4\pi J_ep_0^{-2}(1+p_0^2)^{-0.55},
\end{eqnarray}
where $J_p = 1.9\ \mathrm{cm^{-2}s^{-1}sr^{-1}GeV^{-1}}$, $J_e = 0.02\ \mathrm{cm^{-2}s^{-1}sr^{-1}GeV^{-1}}$, $\beta$ is the proton velocity in units of $c$ and $p_0$ is the particle momentum in $\mathrm{GeV}/c$.
For simplicity, we assume equipartition with the magnetic pressure for the total number densities of the pre-existing CRs \citep[e.g.,][]{1990ApJ...365..544B,2005ARAA..43..337C,2012SSRv..166..307N}, although the CR density and heavy ion abundance in the ISM can be enhanced in regions where a higher concentration of CC SNRs has happened in the past, e.g., OB associations, superbubbles and so on which is beyond the scope of this work.

Ion-neutral damping effects are effective when the shock has decelerated to a point when photo-ionization of the pre-shock medium by the downstream emission becomes partial. The typical shock speed when this happens is $\sim 120\ \mathrm{km/s}$ at which the post-shock temperature has decreased to a few $10^5$~K \citep[e.g.,][]{1989ApJ...342..306H}. Depending on the upstream ionization degree $x$, a spectral break in the accelerated CR spectrum occurs due to the evanescence of the trapping magnetic waves and an enhancement of CR escape above the break momentum. We first calculate the pre-shock ionization fraction \citep{1989ApJ...342..306H} at any given time, which is then used to calculate the local spatial diffusion coefficient and the break momentum in the same way as in \citet{2010ApJ...723L.122U,2011NatCo...2..194M,2015ApJ...806...71L}. The momentum break is then applied to the phase-space distribution of the accelerated particles accordingly. The corresponding equations are 
\begin{eqnarray}
    p_\mathrm{br}/m_pc&=&10B_{-6}^2T_4^{-0.4}n^{-1.5}/(1-x^{0.5})x^{-0.5},\\ \nonumber
    D(x,p) &\propto& \frac{vpc}{3eB(x)}\left(1+{p\over p_\mathrm{br}}\right), \\ \nonumber
    f(x,p>p_\mathrm{br})&=&f_0(x,p)\cdot(p_\mathrm{br}/p).
\end{eqnarray} Figure~\ref{fig:a-fpeDSA} illustrates a result with this feature (see Section~\ref{subsec:res-neutral} for details). The existence of such a momentum break has been suggested recently by gamma-ray observations of older SNRs such as W44 \citep{2011NatCo...2..194M}. 

The non-thermal emission components calculated in this study include inverse-Compton (IC) scatterings, synchrotron radiation, non-thermal bremsstrahlung emission and $\pi^0$-decay \citepalias[][and references therein]{2019ApJ...876...27Y}. 
We do not consider the contribution from secondary particles produced through $\pi^\pm$ decays in this work, which can be important for very dense environments such as giant molecular clouds \citep[see, e.g.,][]{2015ApJ...806...71L} but is out of the scope of this paper. We also focus on the non-thermal emission from particle acceleration at the forward shock and ignore any possible contribution from the reverse shock. 

\subsection{Models for the circumstellar environments and SN ejecta}

\begin{deluxetable*}{l|cccccccc}
\centering
\tablecolumns{8}
\tablewidth{15cm}
\tablecaption{Model parameter} 
\tablehead{
Model & $M_\mathrm{ej}$ & $E_\mathrm{SN}$ & $n_\mathrm{ISM}$ & $B_\mathrm{ISM}$ & $\dot{M}$ & $v_\mathrm{w}$ \\
  & [$M_\odot$] & [$10^{51}\ \mathrm{erg}$] & [cm$^{-3}$] & [$\mathrm{\mu G}$] & [$M_\odot$ yr$^{-1}$] & [km s$^{-1}$]   
}
\startdata
A1\tablenotemark{a} & 1.4 & 1.18 & 10 & 10 & - & - & \\
A2 & 1.4 & 1.18 & 0.1 & 1.0 & - & - & \\
A3 & 1.4 & 1.18 & 10$^{-3}$ & 0.1 & - & - & \\
\tableline
B1\tablenotemark{b} & 12.2 & 1.21 & 0.1 & 1.0 & 1.0$\times$10$^{-4}$ & 20 & \\
B2 & 12.2 & 1.21 & 0.1 & 1.0 & 5.0$\times$10$^{-5}$ & 20 & \\
B3 & 12.2 & 1.21 & 0.1 & 1.0 & 1.0$\times$10$^{-5}$ & 20 & \\
B4 & 12.2 & 1.21 & 0.1 & 1.0 & 5.0$\times$10$^{-6}$ & 20 & \\
B5 & 12.2 & 1.21 & 0.1 & 1.0 & 1.0$\times$10$^{-6}$ & 20 & \\
\enddata
\tablenotetext{a}{All models in group A use an exponential profile for the ejecta, $T_0=10^4\ \mathrm{K}$, $d_\mathrm{SNR}$ = 1.0 kpc, and $\chi_\mathrm{inj}=3.8$.}
\tablenotetext{b}{All models in Group B use a power-law profile for the ejecta with $n_\mathrm{pl}$ = 12, $T_0=10^4\ \mathrm{K}$, $\sigma_\mathrm{w}=0.01$, $d_\mathrm{SNR}$ = 1.0 kpc, $\chi_\mathrm{inj}=3.8$.}
\label{tab:init}
\end{deluxetable*}

\begin{figure}[ht]
    \epsscale{1.15}
    \plotone{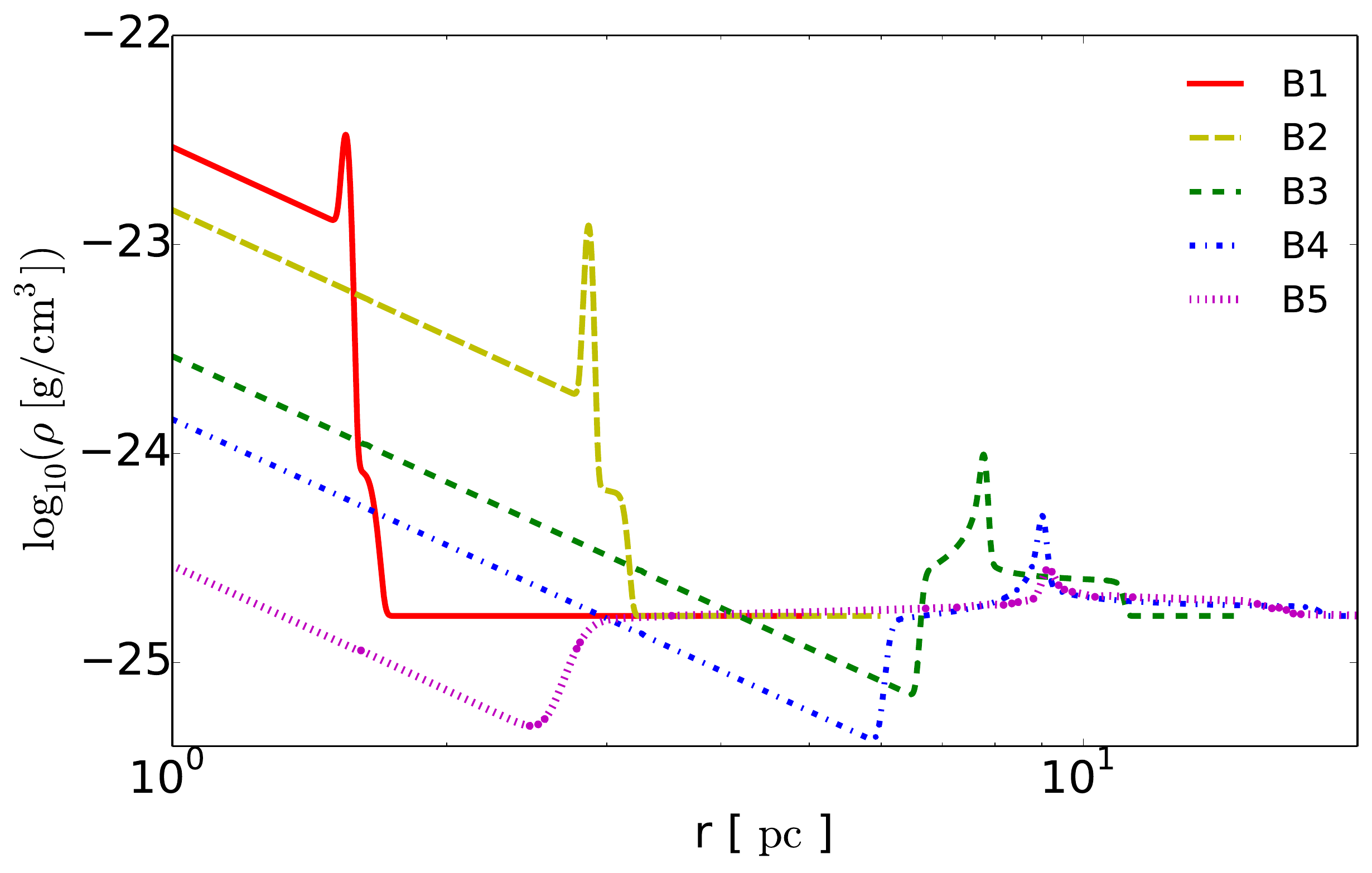}
    \caption{Radial density profiles of the ambient environments in the models of Group B. The red solid, yellow long-dashed, green dashed, blue dash-dotted and magenta dotted lines are associated with models B1, B2, B3, B4 and B5, of which the mass-loss rates of the progenitor are $\dot{M}=10^{-4},5.0\times10^{-5},10^{-5},5.0\times10^{-6}, 10^{-6}\ \mathrm{M_\odot/yr}$, respectively. The total mass-loss is $8$ M$_\odot$ for each model, and the gas density in the outer region is fixed at $n_\mathrm{ISM} = 0.1$~cm$^{-3}$.}
    \label{fig:d-wi}
\end{figure}
\begin{figure}[ht]
    \epsscale{1.15}
    \plotone{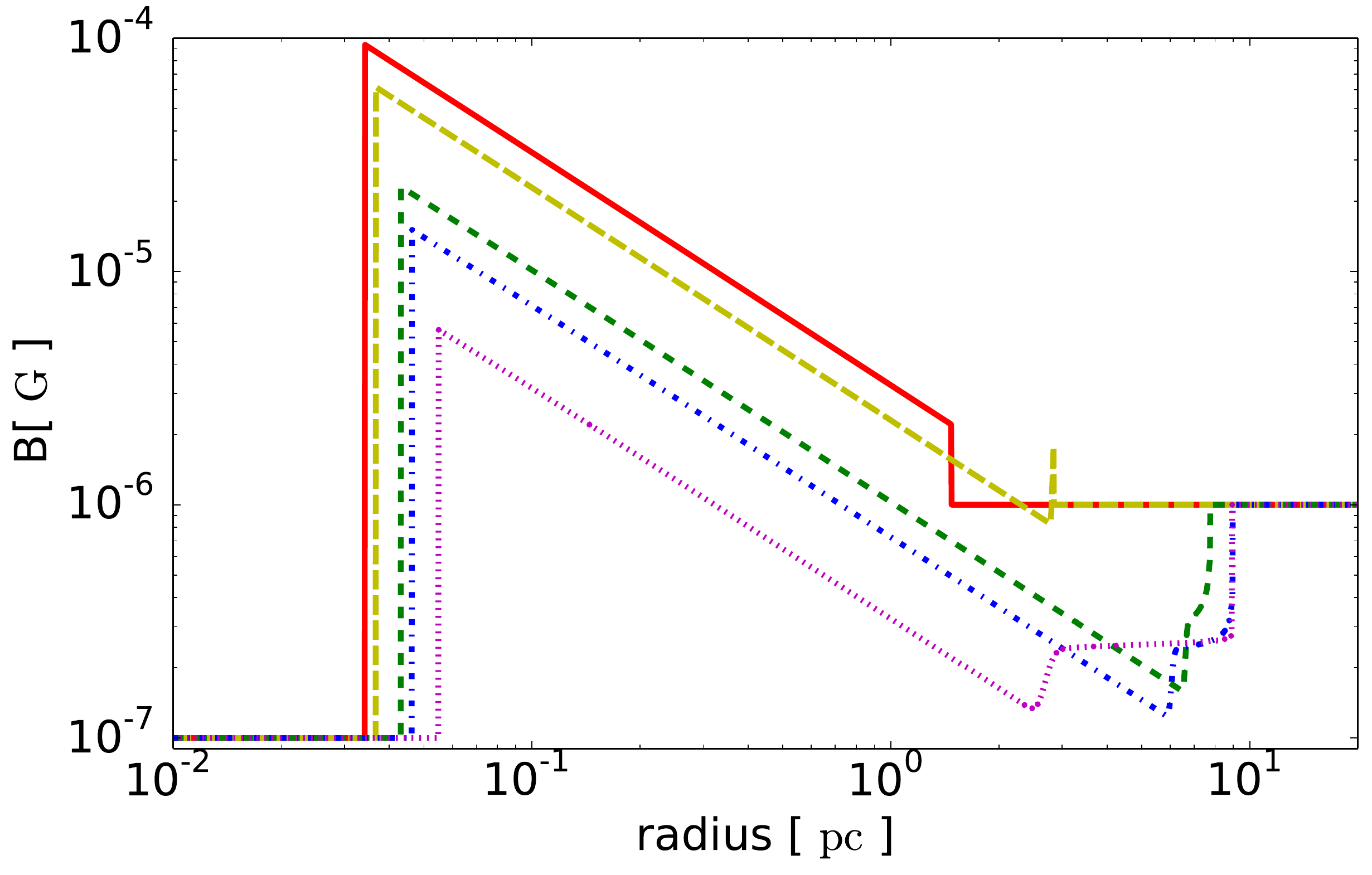}
    \caption{Radial profiles of the magnetic field strength in the ambient environments of the models in Group B. The line formats are the same as Figure~\ref{fig:d-wi}.}
    \label{fig:b-wi}
\end{figure}


We have prepared models in two categories, i.e., Group A (A1 -- A3) and Group B (B1 -- B5), for the circumstellar environments surrounding the SNR. The respective model parameters are summarized in Table~\ref{tab:init}. For the models in Group A, we assume a uniform ambient medium with a constant gas density. This density $n_\mathrm{ISM}$ is varied from $10^{-3}\ \mathrm{cm}^{-3}$ to $10\ \mathrm{cm}^{-3}$. In Group B, we consider the structure created by the progenitor stellar wind with a constant mass-loss rate blowing into a uniform medium. A wind bubble/shell is formed around the ejecta surrounded by a uniform ISM-like gas. 
The CSM structure is obtained by hydrodynamic simulations using the VH-1 code \citep[e.g.,][]{2001ApJ...560..244B} with radiative loss taken into account \citep{1993ApJS...88..253S}. The pre-SN CSM density profiles are plotted in Figure~\ref{fig:d-wi}. The density of the wind deviates from a pure power-law $r^{-2}$ near the interface with the outer ISM whose structure depends on the mass-loss rate.
While the ejecta (progenitor) mass and the pre-SN mass-loss history are related to each other from stellar evolution models, we fix the ejecta mass in this study within each Group and vary the mass-loss rates $\dot{M} = 10^{-6}-10^{-4}\ \mathrm{M_\odot/yr}$ to study the effect of the latter on the SNR evolution. In the free-expanding wind, $\rho(r)=\dot{M}/(4\pi r^2v_w)$ with the wind velocity assumed to be $v_w=20\ \mathrm{km/s}$ for a RSG star. The density of the outer uniform medium is fixed at $n_\mathrm{ISM}=0.1\ \mathrm{cm}^{-3}$. We assume that the CSM is composed of a RSG wind and ignore any mass loss from the main-sequence (MS) and other possible mass loss phases. We recognize that the MS stellar wind prior to the RSG phase can impose a large influence on the SNR evolution which can alter the light curves/spectral evolution in a non-negligible way, as shown by a number of previous works which investigated models taking into account the mass loss in the MS phase and their interactions with the subsequent RSG wind and in some cases (e.g., for a Type Ib/c progenitor) Wolf-Rayet wind and binary mass transfer as well \citep[e.g.,][and reference therein]{2021ApJ...919L..16Y,Yasuda_2022,2022AA...661A.128D}. We are ignoring the MS wind bubble and for that matter episodic mass loss for simplicity here to focus on the systematic effect of $\dot{M}$ on the long-term emission evolution and leave the discussion on the MS wind effect to a future work. 

The initial magnetic field strength profiles are plotted in Figure~\ref{fig:b-wi}. There are “jumps" in the magnetic field strength at the interface between the wind and the ISM in our models, which are also featured in \citet{2022ApJ...926..140S}. Stemming from this jump, we have confirmed a ``double-bump" feature in the gamma-ray SED (via IC and bremsstrahlung) resulted when the shock propagates through the interface (see Section~\ref{subsec:res-csm}), which is also observed in \citet{2021ApJ...919L..16Y,Yasuda_2022,2022ApJ...926..140S}. The magnetic field strength in the wind is determined by the magnetization parameter $\sigma_\mathrm{w}\equiv (B^2/8\pi)/(\rho v_\mathrm{w}^2/2) = 10^{-2}$ \citep[e.g.,][]{LEN2012,2013ApJ...767...20L}. The magnetic field strength in the ISM-like ambient medium for models in both Groups A and B are on the other hand determined by a scaling proportional to $\sqrt{n_\mathrm{ISM}}$ assuming magnetic flux freezing under isothermal condition \citep[e.g.,][]{1999ApJ...520..706C,2010ApJ...723L.122U}. For $n_\mathrm{ISM}=0.1\ \mathrm{cm}^{-3}$ and $T_\mathrm{ISM}=10^4\ \mathrm{K}$, we assume $B_\mathrm{ISM}=1\ \mathrm{\mu G}$. 

We can further categorize the initial CSM profiles into two types: B1--B2 and B3--B5. Models B1 and B2 have relatively large mass-loss rates which result in a dense wind shell whose spatial scale is mainly dictated by the mass loss duration prior to explosion. Models B3--B5 on the other hand form a wind ``bubble'' surrounded by a dense shell whose dynamics is determined by mechanical (pressure) balance instead. B5 in particular has a relatively small cavity-like structure due to the low mass-loss rate and hence gas ram pressure. The total mass-loss is fixed at $8~\mathrm{M_\odot}$ for all models in Group B (see below). These differences in the CSM profiles will reflect strongly in the resulted light curves in the SNR phase.  


We assume an ejecta with energetics typical of a Type Ia SN for Group A, and an ejecta from the core collapse (CC) explosion of a RSG star for Group B.
We use the \textit{DDT12} model \citep[][and references therein]{2018ApJ...865..151M} for the Type Ia ejecta which is representative of a ``normal'' thermonuclear explosion of a near-Chandrasekhar mass white dwarf star, i.e., $M_\mathrm{ej}=1.4$~M$_\odot$, $E_\mathrm{SN}=1.18\times10^{51}\ \mathrm{erg}$ with an exponential profile \citep{1998ApJ...497..807D}. A RSG model \textit{s25D} \citep{2015ApJ...803..101P,2010ApJ...724..341H} is used for the CC SNRs in Group B with an original ZAMS mass of 25~M$_\odot$, for which $M_\mathrm{ej}=12.2$~M$_\odot$, $E_\mathrm{SN}=1.21\times10^{51}\ \mathrm{erg}$ with a power-law envelope model \citep{1999ApJS..120..299T} whose index is $n_\mathrm{pl}=12$ \citep{1999ApJ...510..379M} for the ejecta density profile. This model involves a total mass-loss of 8~M$_\odot$ through stellar wind prior to CC. 

\section{Results and Discussion} \label{sec:results}

\subsection{Models with a uniform medium}\label{subsec:res-uni}

\begin{figure}[ht]
    \gridline{\fig{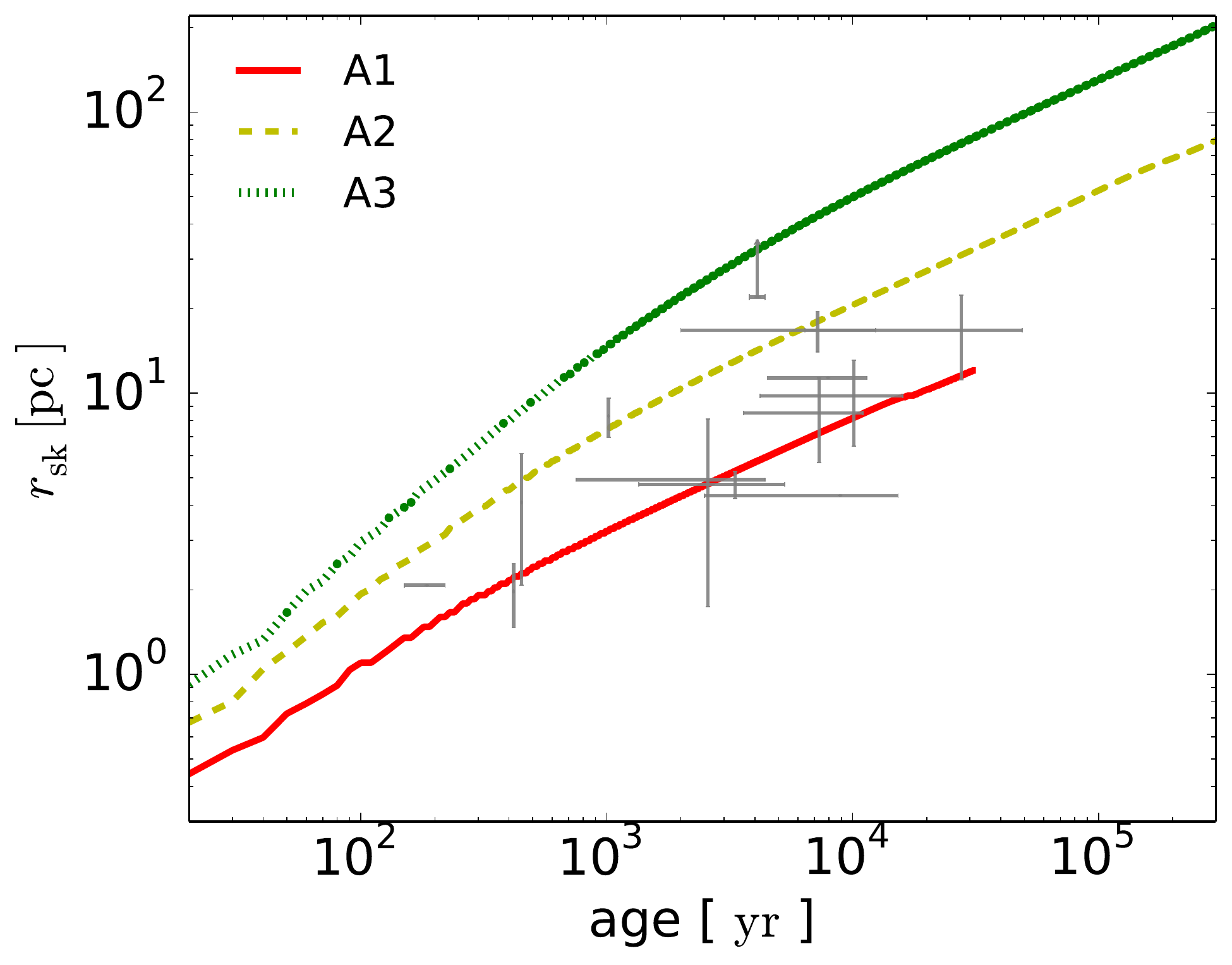}{0.4\textwidth}{(a) Shock radius}}
    \gridline{\fig{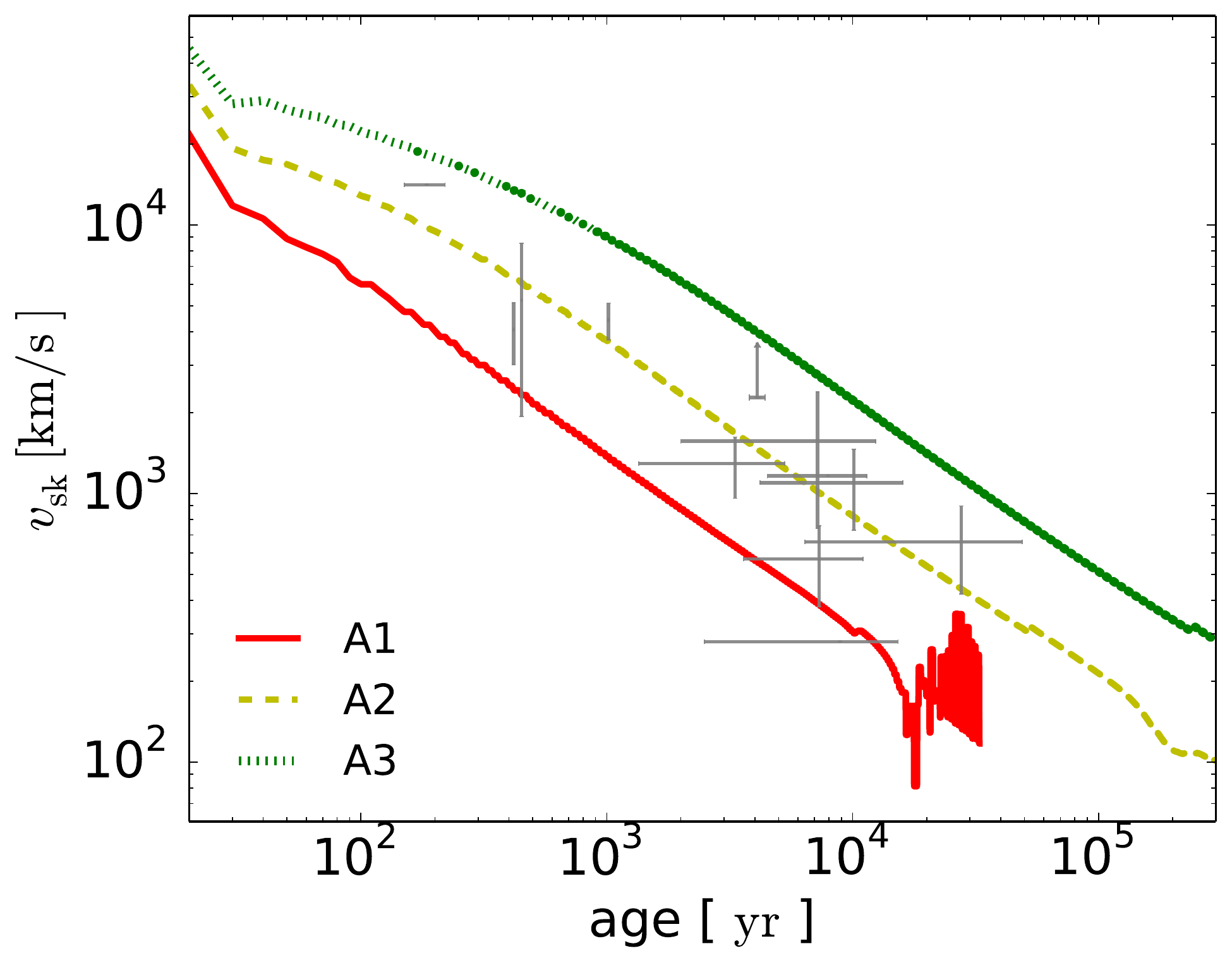}{0.4\textwidth}{(b) Shock velocity}}
    \gridline{\fig{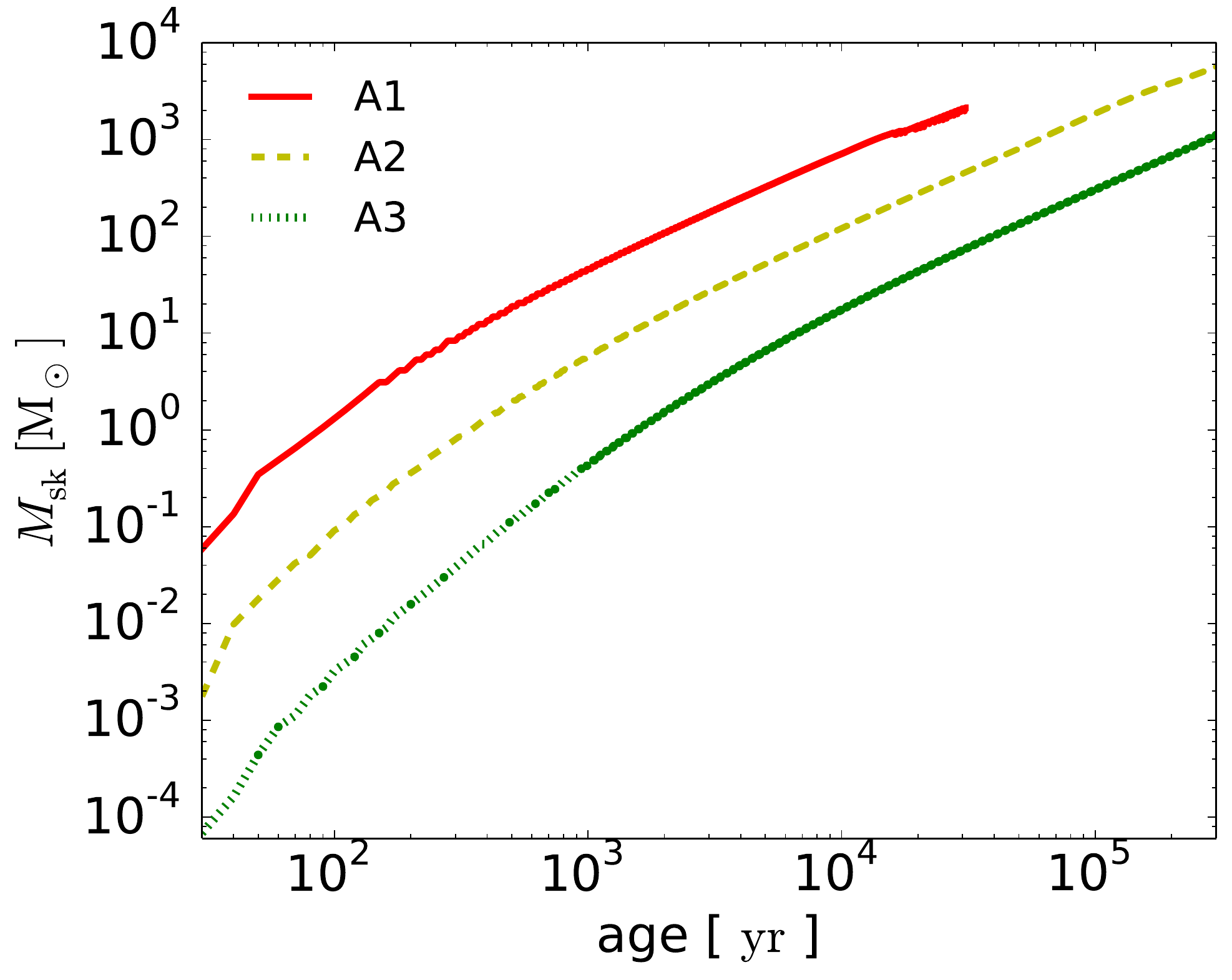}{0.4\textwidth}{(c) Shocked mass}}
    \caption{Shock radius (a), velocity (b) and shocked gas mass (c) as a function of age for Group A compared with the observation data of Type Ia SNRs shown by the data points. The red solid, yellow dashed and green dotted lines show the model results for models A1, A2, A3 of which the ISM densities are $n_\mathrm{ISM}=10, 0.1, 10^{-3}\ M_\odot/\mathrm{yr}$, respectively.}
        \label{fig:a-rv}
\end{figure}

\begin{figure*}[ht]
	\centering	
	\plotone{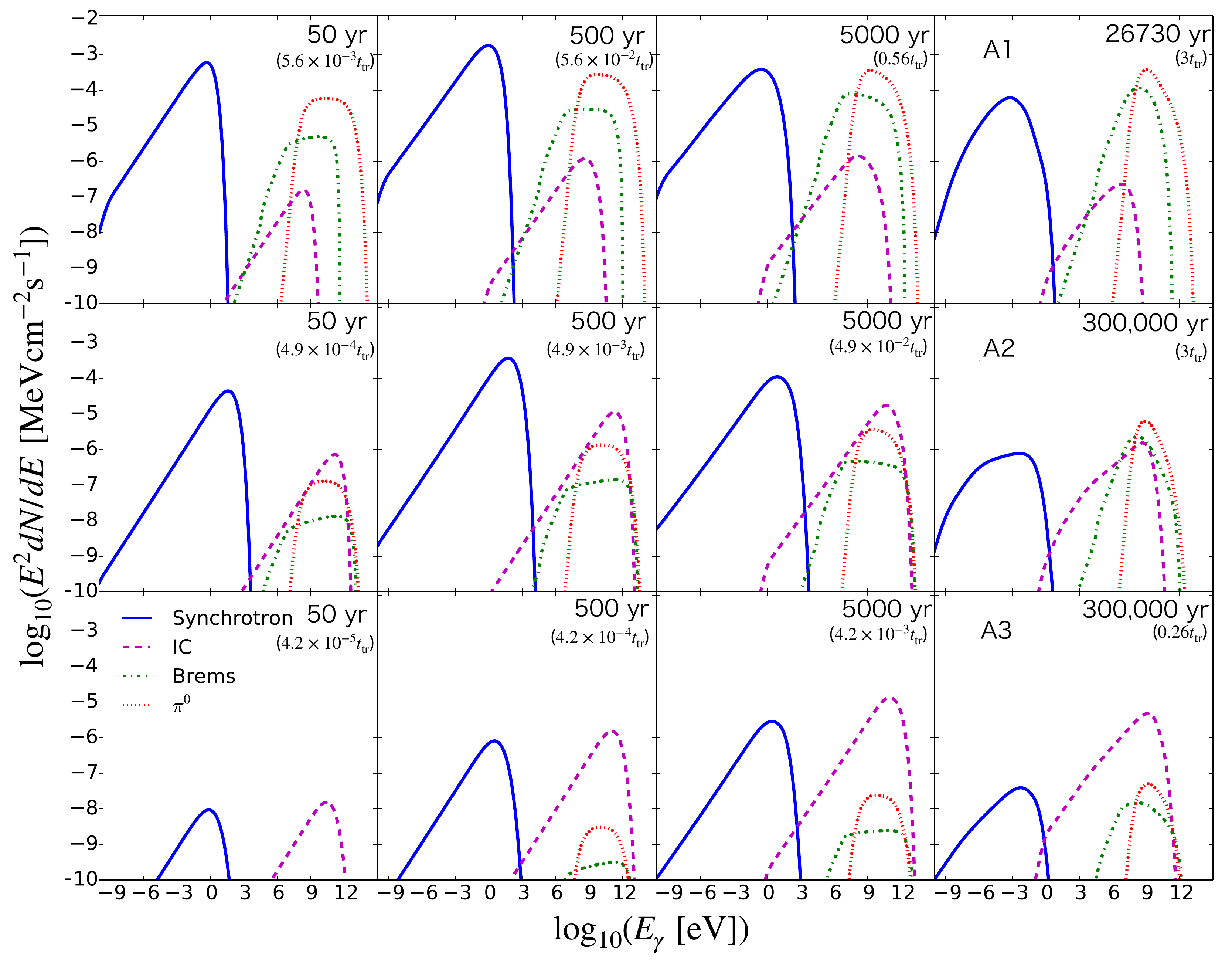}
    \caption{Time evolution of the broadband non-thermal spectra from models in Group A for each emission component until the radiative phase. Snapshots at t = 50 yr, 500 yr, 5000 yr, and $3t_\mathrm{tr}$ (or $3 \times 10^5$ yr for model A3) are shown from left to right, and models A1, A2 and A3 ($n_\mathrm{ISM}=10\ \mathrm{cm}^{-3}, 0.1\ \mathrm{cm}^{-3}, 10^{-3}\ \mathrm{cm}^{-3}$ respectively) from top to bottom. The emission components include synchrotron (blue solid), non-thermal bremsstrahlung (green dash-dotted), inverse-Compton (magenta dashed) and $\pi^0$-decay (red dotted). }
	\label{fig:a-sed}
 \end{figure*}

\begin{figure}[p]
    \gridline{\fig{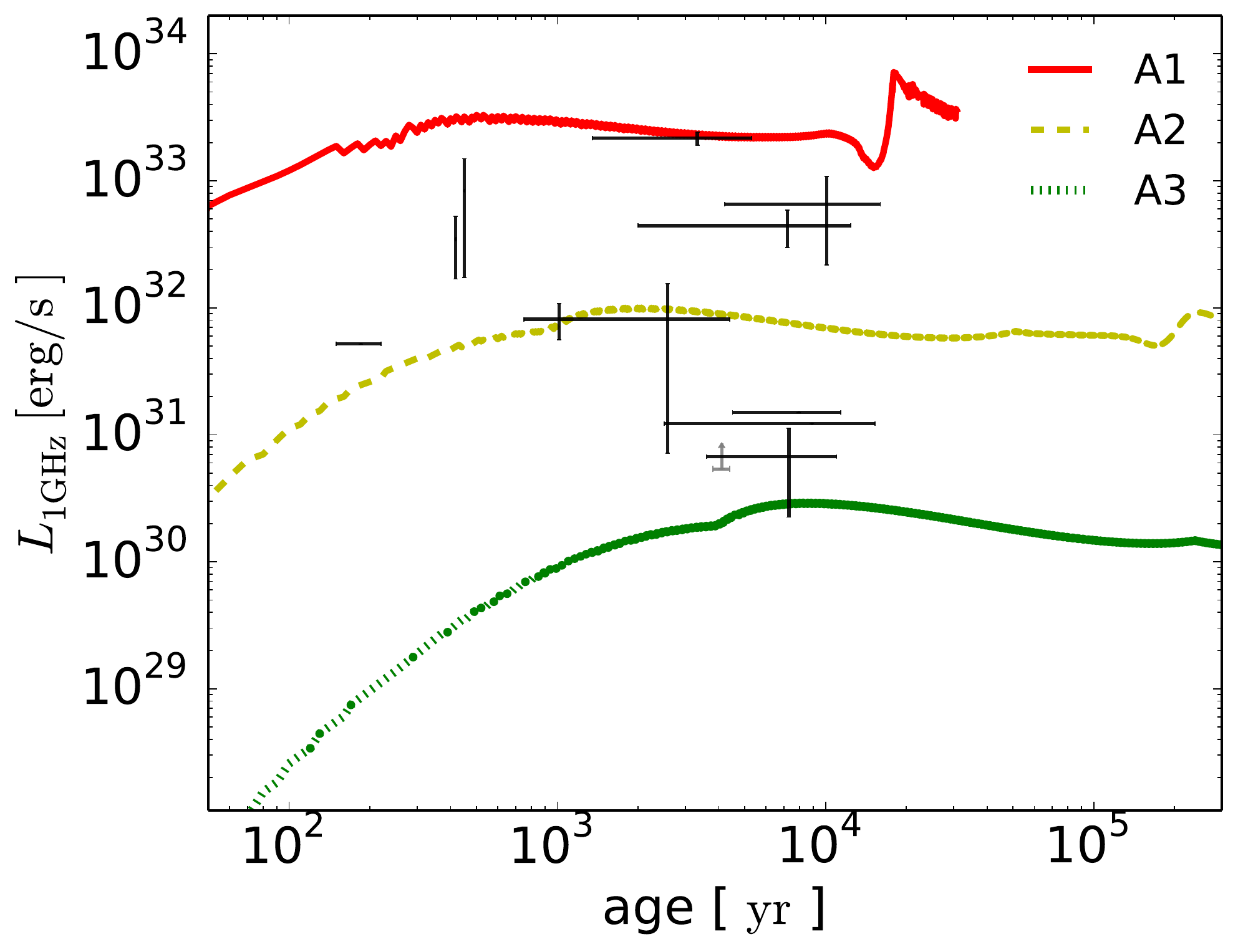}{0.4\textwidth}{(a) 1 GHz luminosity}}
    \gridline{\fig{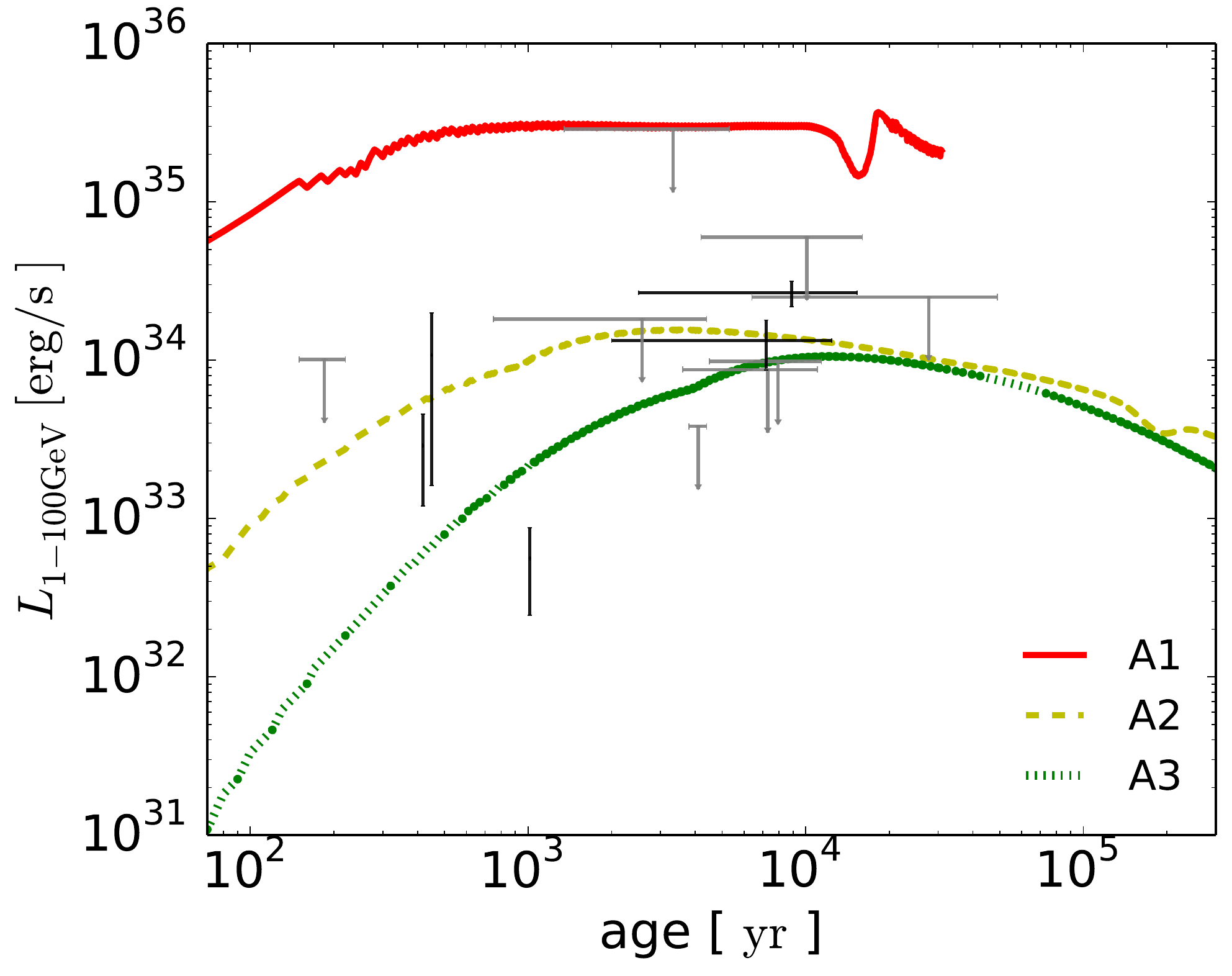}{0.4\textwidth}{(b) 1--100 GeV luminosity}}
    \gridline{\fig{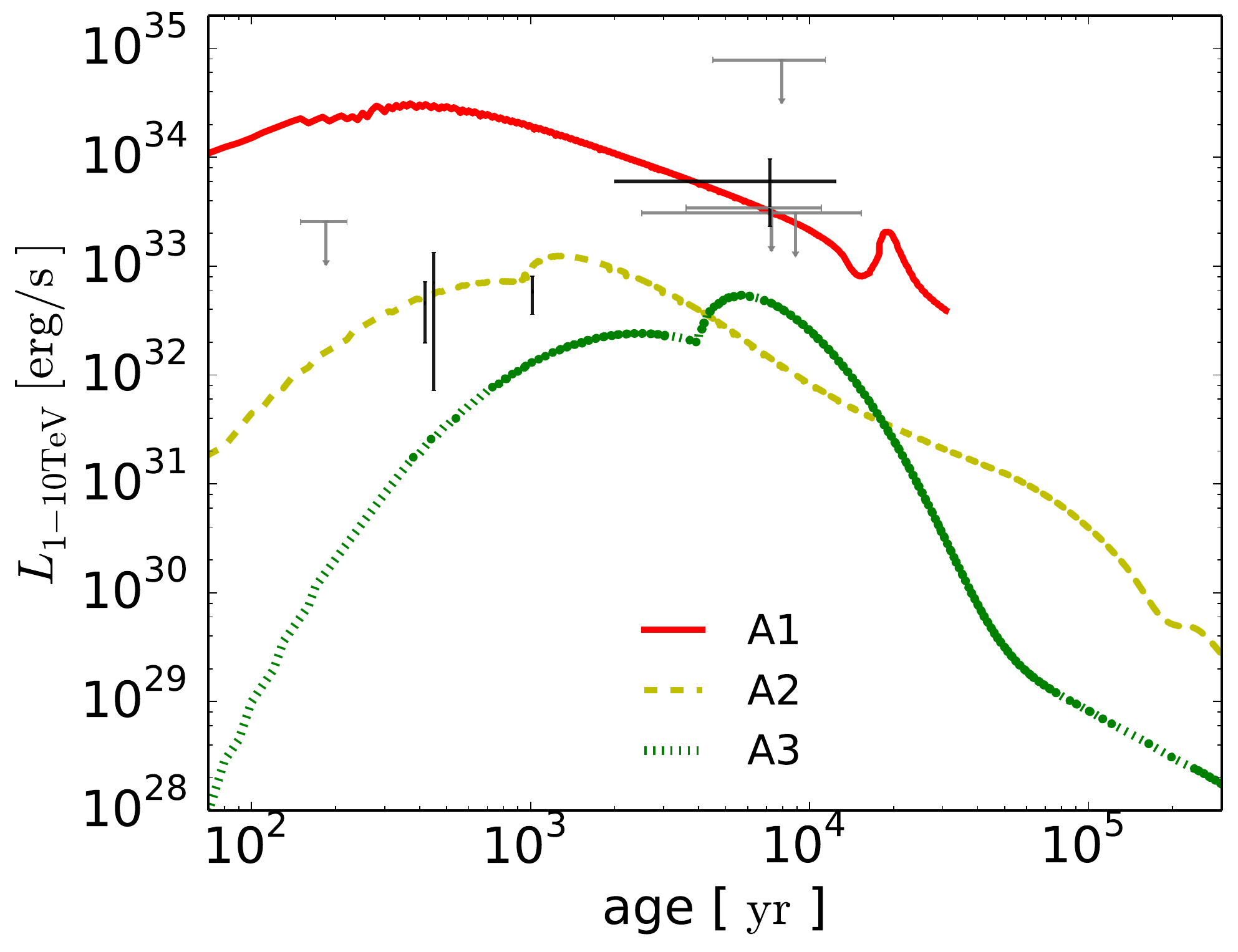}{0.4\textwidth}{(c) 1--10 TeV luminosity}}
    \caption{Time evolution of the luminosities in three energy bands for models in Group A compared to the observation data of Type Ia SNRs (data points). The line formats are the same as in Figure~\ref{fig:a-rv}. Panel (a) shows the luminosity at 1 GHz, (b) from 1 GeV to 100 GeV and (c) from 1 TeV to 10 TeV.}
        \label{fig:a-lc}
\end{figure}

  \begin{figure}[p]
    \gridline{\fig{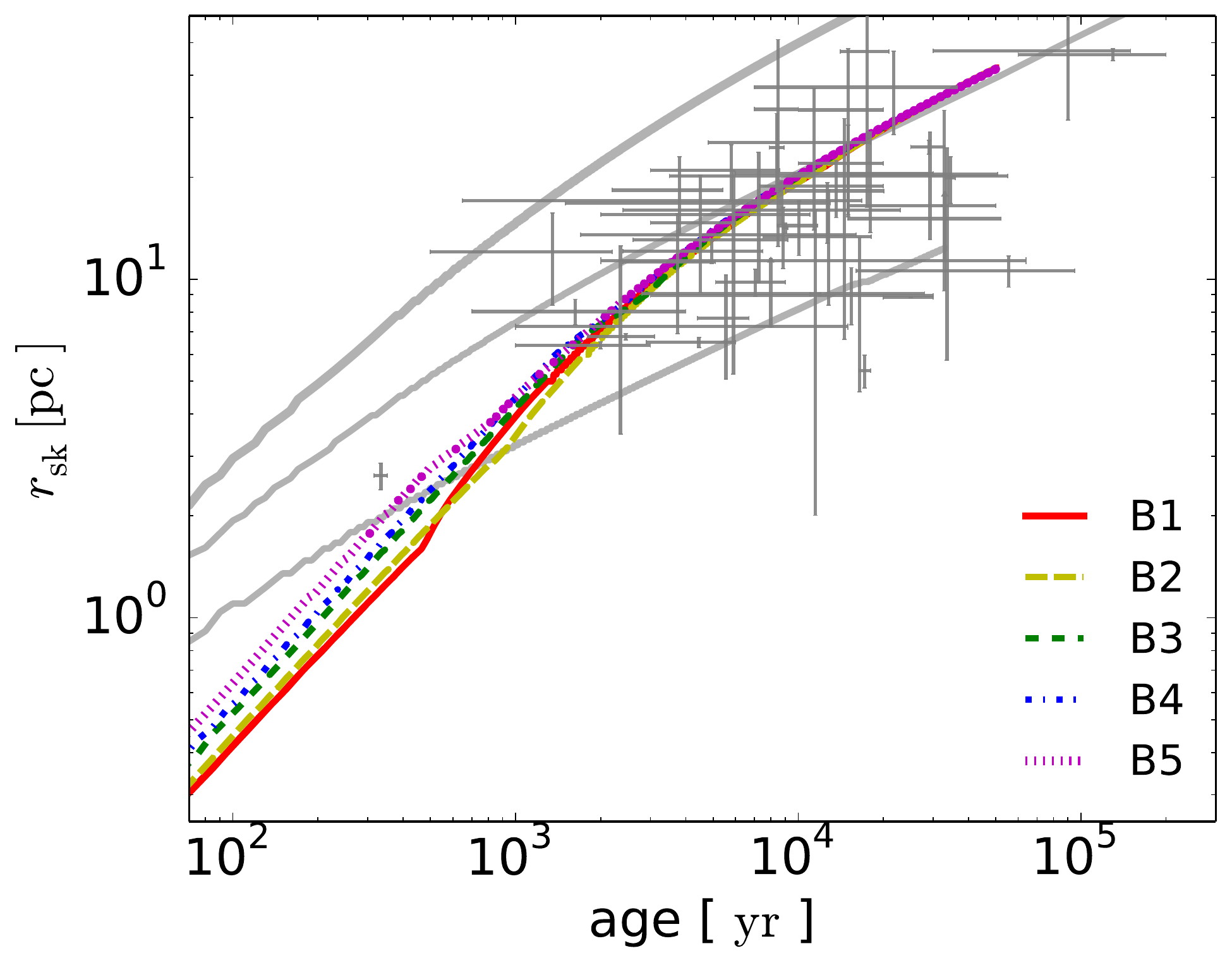}{0.4\textwidth}{(a) Shock radius}}
    \gridline{\fig{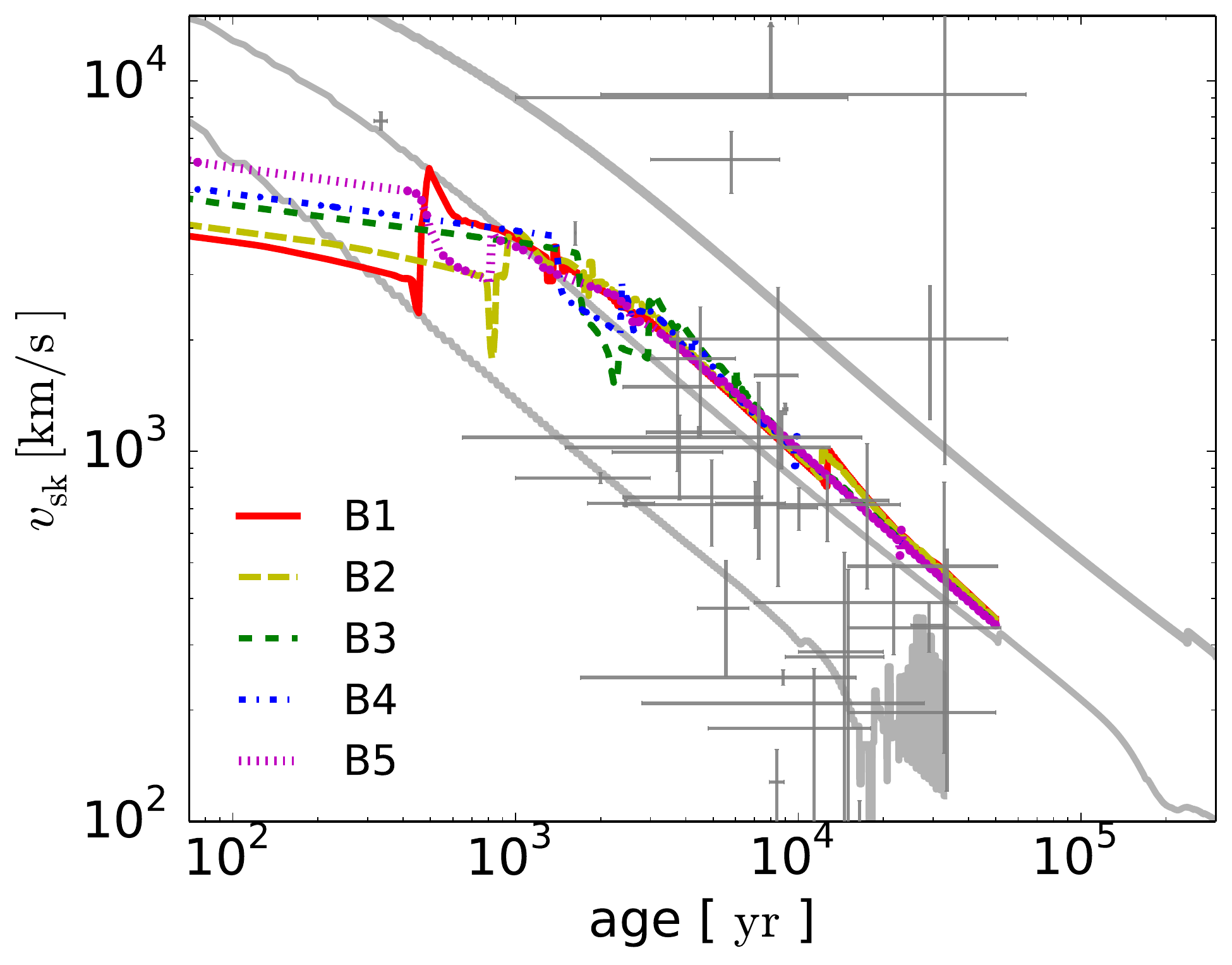}{0.4\textwidth}{(b) Shock velocity}}
    \gridline{\fig{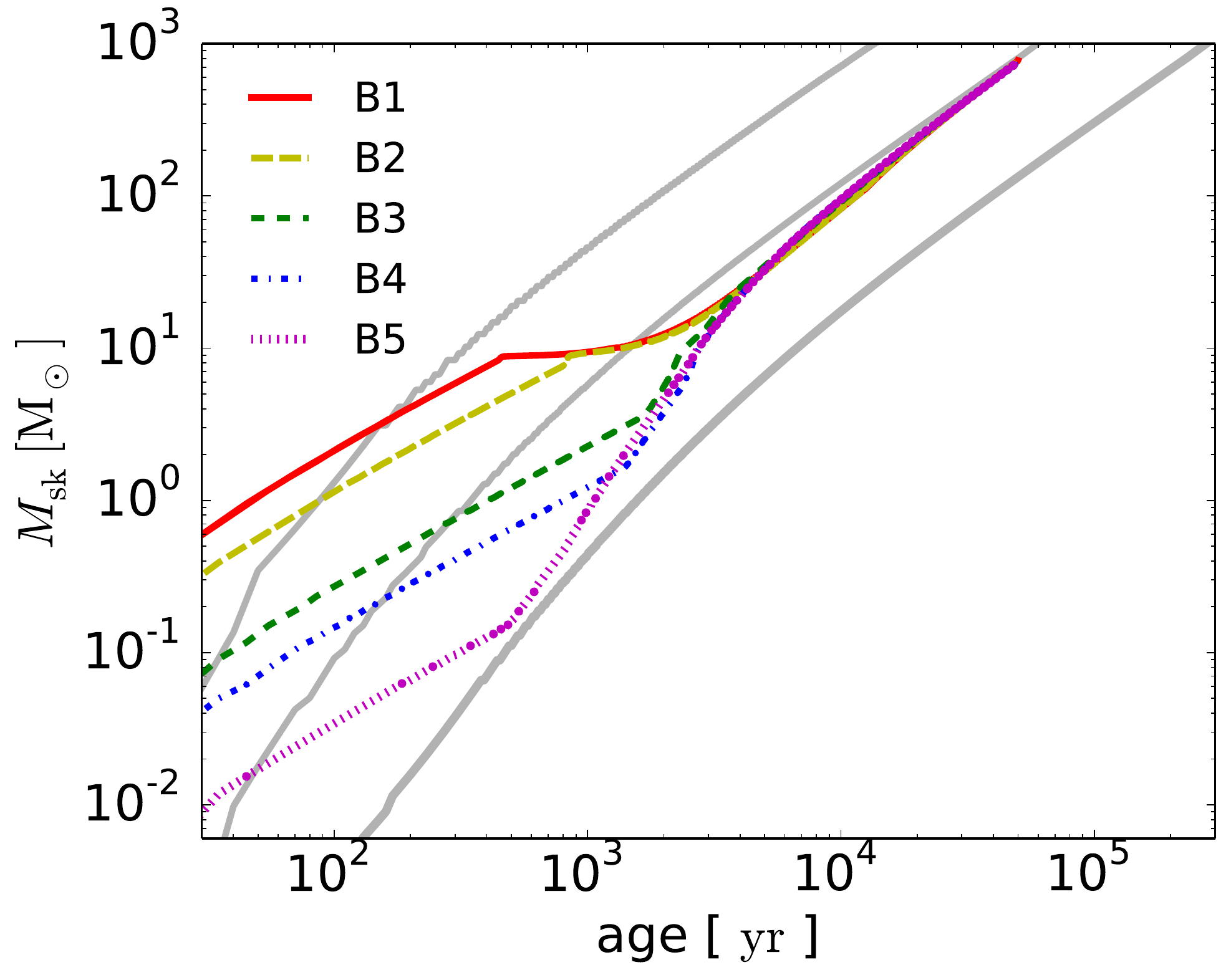}{0.4\textwidth}{(c) Shocked mass}}
    \caption{Same as Figure~\ref{fig:a-rv} but for Group B, overlaid with the results from models in Group A in grey lines and CC SNRs observation data for reference. The line formats are the same as in Figure~\ref{fig:d-wi}.}
        \label{fig:b-rv}
\end{figure}

In this Section, we first elaborate on the results from the models in Group A with a uniform ambient medium, which will also serve as a reference for the discussion of the models in Group B in which more complicated CSM environments are involved.  
Figure~\ref{fig:a-rv} shows the time evolution of the shock radius $r_\mathrm{sk}$, velocity $v_\mathrm{sk}$ and shocked mass $M_\mathrm{sk}$. With the same explosion energetics of a typical Type-Ia SN for the three models, a lower density ISM leads naturally to a larger remnant and a faster blastwave at any given age. A higher ISM density such as in model A1 also results in an earlier transition to the radiative phase as the shock speed has decreased to $\sim$ 100 km/s. This can be witnessed from the oscillation of the shock speed after the transition commences, which comes from the interaction of the newly formed post-shock cold dense shell with the fast expanding gas in the interior \citep[see, e.g.,][]{2015ApJ...806...71L} and is more prominent for a higher $n_\mathrm{ISM}$ where the denser radiative shell formed behind the shock imposes a larger influence on the bulk hydrodynamic evolution. This oscillation however is known to be exaggerated by 1D-treatments and is expected to be much milder in multi-dimensional simulations in which the spherical symmetry is broken \citep{2021MNRAS.505..755P}. For a sanity check, data points are compiled from measurements of known Type Ia SNRs for reference listed in Table~\ref{tab:obs}, which show a general agreement with the range of results yielded by our models.     


Figure~\ref{fig:a-sed} shows the broadband SED evolution for the non-thermal emission. We first confirm that the results before an age of $5,000$ yrs are in broad agreement with those reported in \citetalias{2019ApJ...876...27Y}. For example, we see a similar dependence of the hadronic versus leptonic origin of the gamma-rays on the ISM density, as well as its variation with the SNR age. The subsequent SED evolution beyond the Sedov phase is first explored in this work. At $t = 3 t_{tr}$, the remnants in model A1 and A2 have already entered the radiative phase. We can see a significant softening of the spectra across the entire frequency band. This can be attributed to the now decelerated shock with a velocity $\sim 100 - 200$~km/s, at which the velocity of the magnetic scattering centers in the upstream can no longer be neglected. Both the maximum attainable energy and the effective compression ratio felt by the accelerating CRs decrease, resulting in a soft CR spectrum. The spectral shape of the synchrotron and IC components are remarkably different between A1 and A2 (i.e., a stronger radio and weaker IC contributions and a lower energy cutoff in model A1), which can be explained by the higher averaged magnetic field strength in the shocked plasma in model A1 with a higher ISM density and hence a faster synchrotron loss for the electrons. The dense cold shell formed behind the shock in model A1 in the radiative phase also contributes to an amplification of the magnetic field and gas density due to the fast compression during the formation of the shell. The GeV emission from pion-decay as well as the bremsstrahlung contribution in the hard X-ray and MeV energy range (whose luminosities are proportional to $n_\mathrm{ISM}^2$) are also much more prominent in A1, resulting in an interesting distinction in spectral shape with model A2. Model A3 on the other hand shows a relatively monotonic evolution in comparison which is mainly dominated by the fast adiabatic expansion of the SNR in a tenuous medium. Over the course to $3\times10^5$ yrs, radiative cooling never plays an important role for such a low ISM density.  

One of the novel features our models have discovered is that we cannot confirm the emergence of a clear spectral break in a radiative SNR (models A1 and A2) which is expected from the effect of ion-neutral damping of the magnetic waves in a partially ionized shock precursor \footnote{In \citet{2020AA...634A..59B}, this break feature is explained by a fast CR escape instead.}. In contrast to, e.g., \citet{2015ApJ...806...71L} who only considered the local emission from a cloud shock, the difference comes from the fact that we initialize our simulations from the SN explosion so that the contribution from all CRs accelerated by the SNR shock before the SNR becomes radiative cannot be neglected. As a result, the contribution from the CRs accelerated in the radiative phase is only partial to the overall volume-integrated SED, making the break feature much less visible. However, we note that a momentum break does indeed appear in the local CR spectra immediately behind the radiative shock (see Section~\ref{subsec:res-neutral} for details). 

Figure~\ref{fig:a-lc} shows the light curves predicted by our models in three energy bands: 1 GHz in Panel (a), gamma-ray integrated over the range of 1--100 GeV in Panel (b) and over the range of 1--10 TeV in Panel (c). 
The time variation of the radio (synchrotron) emission can be explained by the balance between the contribution from the newly injected CRs in accumulating shocked ISM and the decrease in density of the CRs  accelerated in the past and advected downstream. The latter suffer from adiabatic cooling, and the magnetic field density also declines along with the expansion of the SNR. In model A1 with a high ambient density, the luminosity increases rapidly in the first few 10 yrs as the shocked ISM mass quickly accumulates. After around several $10^2$ yr, the luminosity saturates and begins to decline due to adiabatic cooling. We can see a luminosity boost after $10^4$ yr by a factor of $\sim$ 4, which comes from the contribution from the formation of a thin cold dense shell behind the radiative shock and the resulted compression of the CR, gas and magnetic field densities. Models A2 and A3 show similar behaviors to model A1 despite that the luminosities saturate and decrease at later ages, i.e., $\sim 10^3$ yr (A2), $\sim 10^4$ yr (A3) proportional to their Sedov ages, reflecting the differences in their evolution (i.e., shock velocity and total mass of shocked gas). The luminosity boost in model A2 is much milder than what is observed in model A1 since the radiative gas shell is much less prominent. 
We also overlay observation data from Table~\ref{tab:obs} onto the light curves for comparison. Our results are found to be consistent with the observed Type Ia SNRs for an ambient density ranging from $10^{-3}-10\ \mathrm{cm}^{-3}$. To this end, further constraints on parameters for the individual observed remnants such as their magnetic field strengths can be obtained from detailed spectral modeling for each object including the X-rays and gamma-rays, but is beyond the scope of this work. 

While the overall trends are found to be qualitatively similar between the gamma-rays and radio emission, our results predict similar luminosities for models A2 and A3 in the gamma-rays after $\sim10^4$ yr despite the different ambient densities and total mass in the shocked gas involved. 
The SNRs in both A2 and A3 emit gamma-rays mainly through the IC channel up to $10^4$ yrs of evolution (albeit with a more appreciable mix from $\pi^0$-decay in A2 for obvious reasons). The shocked masses at that point should then roughly differ between the two by a factor of $(n_{0,\mathrm{A2}}/n_{0,\mathrm{A3}})^{2/5} \sim 6$ from the standard Sedov solution, which can be confirmed in Figure~\ref{fig:a-rv}(c). From that, the additional effect of a stronger spectral steepening experienced by the CRs in A2 then compensate for this factor of a few and bring the luminosities of A2 and A3 close to each other. Indeed, at an age $\sim10^4$ yrs, the shock in model A2 has decelerated to a point such that the Alfv\'enic Mach number $M_\mathrm{A}$ has decreased to a few, resulting in a steeper CR spectrum than in model A3.
The hadronic emission does suffer from spectral steepening as well for the same reason above, but the slower SNR expansion in the denser medium and the long energy loss timescale from pion production for the protons ensure that the hadronic gamma-rays do not decline rapidly with time. The extra multiplicative factor of $\rho$ in the normalization scaling of the $\pi^0$-decay emission is also responsible for the boost of the gamma-ray luminosity in the hadronic dominated model A1 against A2.

The light curve in the TeV band is additionally affected by the evolution of the maximum CR momenta which in turn determine the gamma-ray spectral cutoff energies. An abrupt increase in the luminosity can be observed at certain ages, especially in the TeV band for model A2 and A3. This comes from the non-linear DSA effects kicking in as the SNR enters the Sedov phase and the shock velocity decreases to a few $10^3$~km/s, which results in a non-linear increase in the DSA efficiency and an accompanying amplification of the magnetic field and increase of the maximum energy of the CRs. The TeV luminosity boost in model A3 is especially strong due to the smaller downstream magnetic field and hence a less influence from synchrotron loss on the gamma-ray (mainly from IC here) cutoff energy compared to other models. 
As the shock velocity decreases further, the non-linear DSA effects subside, and the luminosity levels converge back to those expected in the test-particle DSA limit. Meanwhile, the shock compression ratios in our models are generally suppressed compare to those usually expected from an efficient NLDSA. One reason is from the inclusion of the magnetic dynamical feedback effect as described in \citet{CBAV2009a} which makes the fluid less compressible. Moreover, the Alfv\'enic-drift model has an effect of spectral steepening (softening) on the accelerated CRs, such that the pressure from the counter-streaming CR on the inflowing gas is further reduced in the shock precursor. These two factors combined lead to a reduction in the shock compression ratio and prevent a strong shock modification due to the non-linear CR feedback as seen in some other works. 
%
%
Compared to the gamma-ray observation data (with upper limits) from Table~\ref{tab:obs}, our simulation results are also found to be in bulk agreement. The comparison suggests that most observed Ia SNRs are interacting with an ambient medium with densities $\lesssim 0.1\ \mathrm{cm}^{-3}$. SN 1006 (1016 yrs old) at a high Galactic latitude is known to be interacting with a tenuous ISM which is indeed suggested to be the case by our models as well. 
The apparent discrepancy for the object RCW 86 (2000--12400 yrs old) which is known to interact with a dense molecular cloud \citep{2017JHEAp..15....1S} can be possibly due to a deviation from a simple uniform ISM-like environment encountered by the SNR. 

\subsection{Models with stellar winds}\label{subsec:res-csm}

  \begin{figure*}[ht]
	\centering	
	\plotone{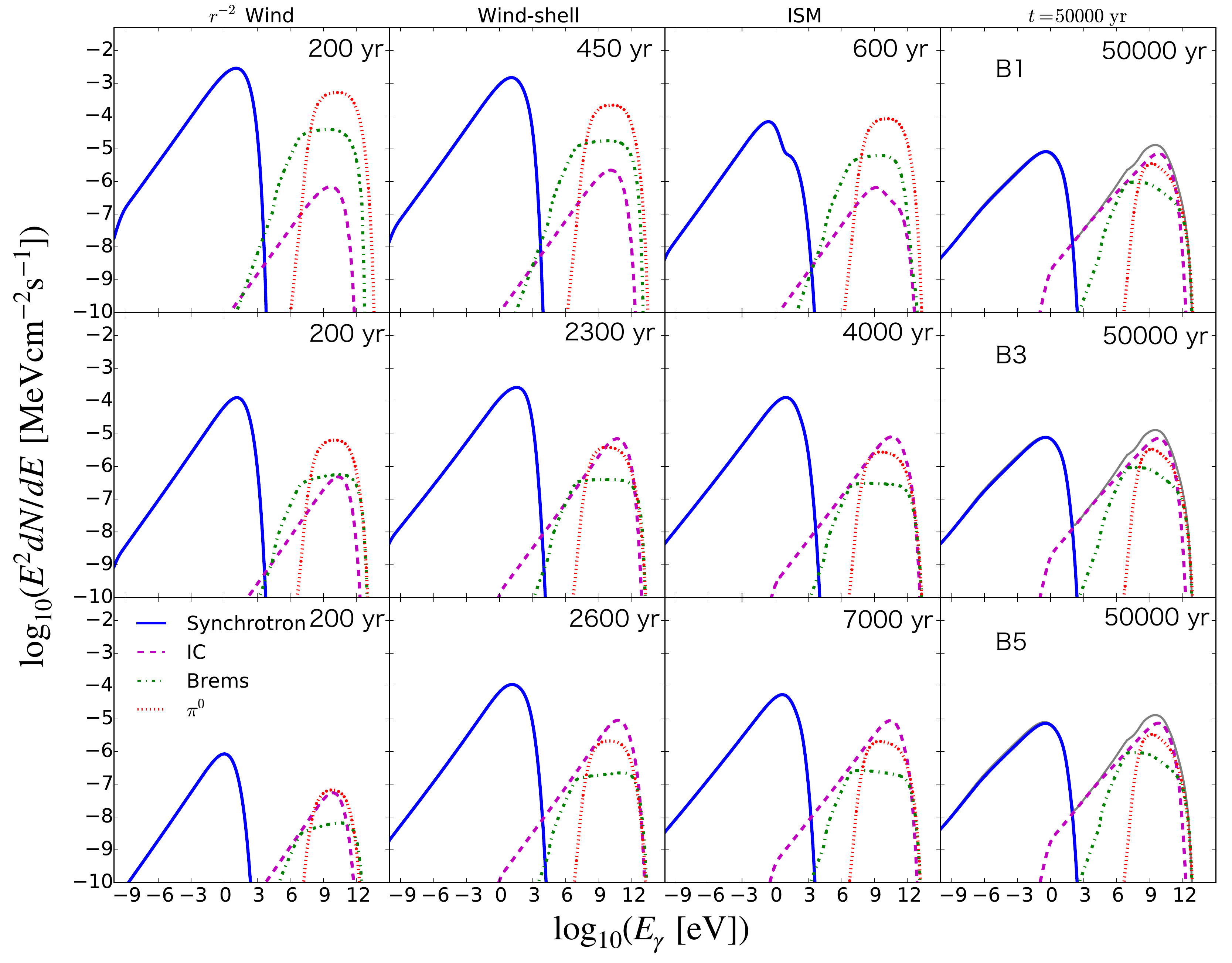}
    \caption{The same as Figure~\ref{fig:a-sed} but for the models in Group B. $\dot{M}=10^{-4}, 10^{-5}, 10^{-6}\ \mathrm{M_\odot/yr}$ are shown from the top to the bottom which are for models B1, B3, B5 respectively. The ages plotted are 200, 450, 600, 50000 yr in B1, 200, 2300, 4000, 50000 yr in B3, and 200, 2600, 7000, 50000 yr in B5. Grey lines in the most right panel show the total component in model A2 in the age of 50000 yr.}
	\label{fig:b-sed}
 \end{figure*}

 \begin{figure}[ht]
    \gridline{\fig{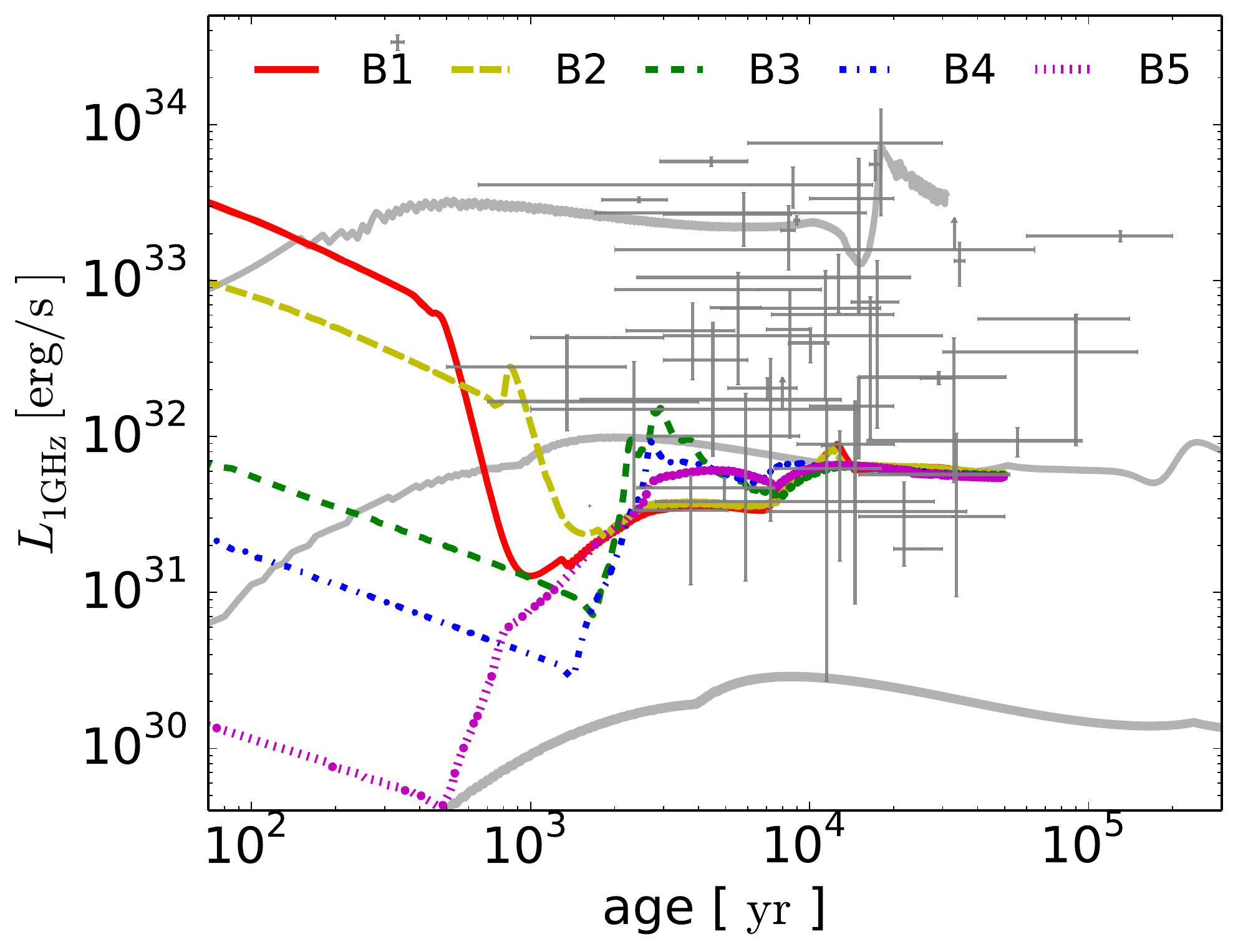}{0.4\textwidth}{(a) 1 GHz luminosity}}
    \gridline{\fig{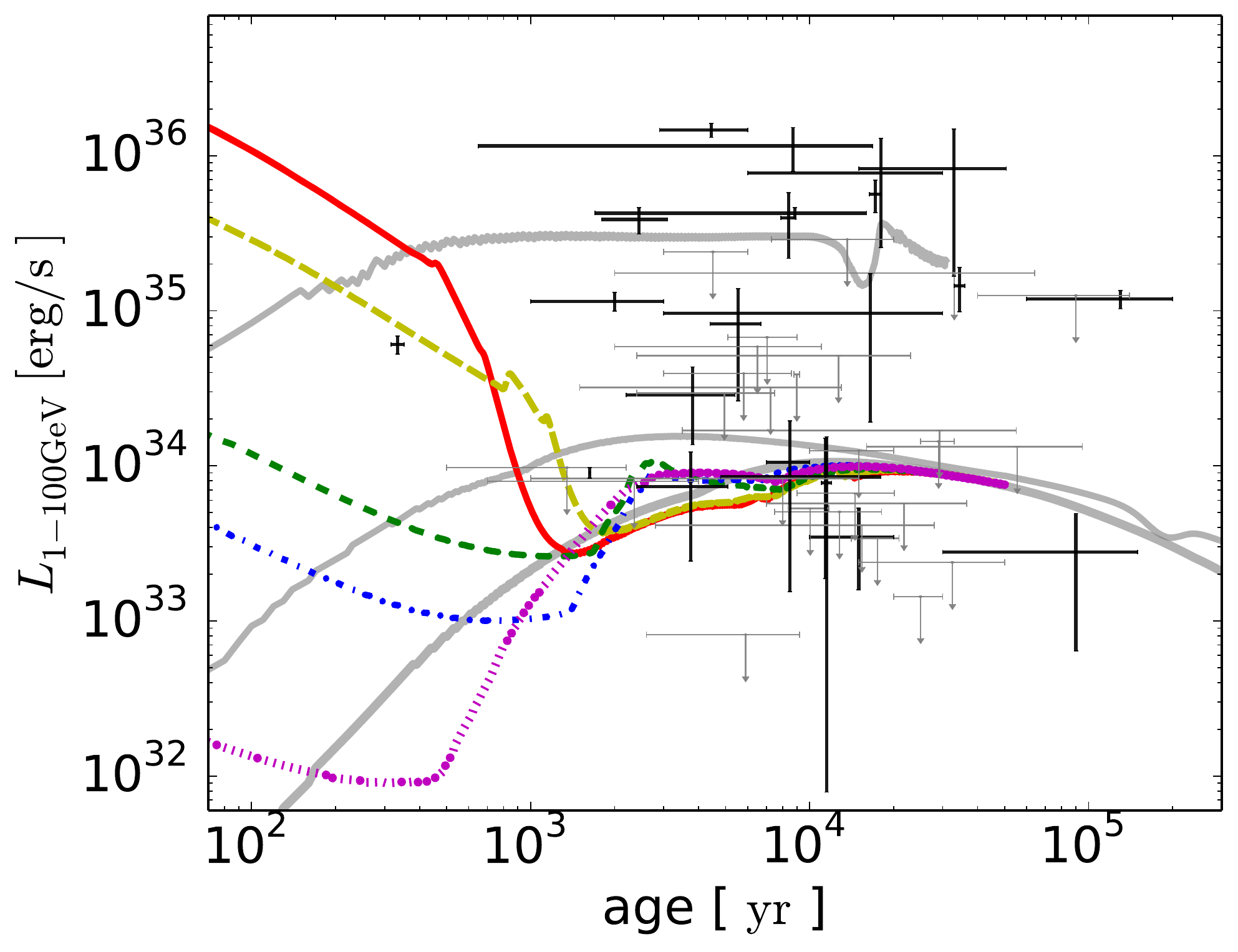}{0.4\textwidth}{(b) 1--100 GeV luminosity}}
    \gridline{\fig{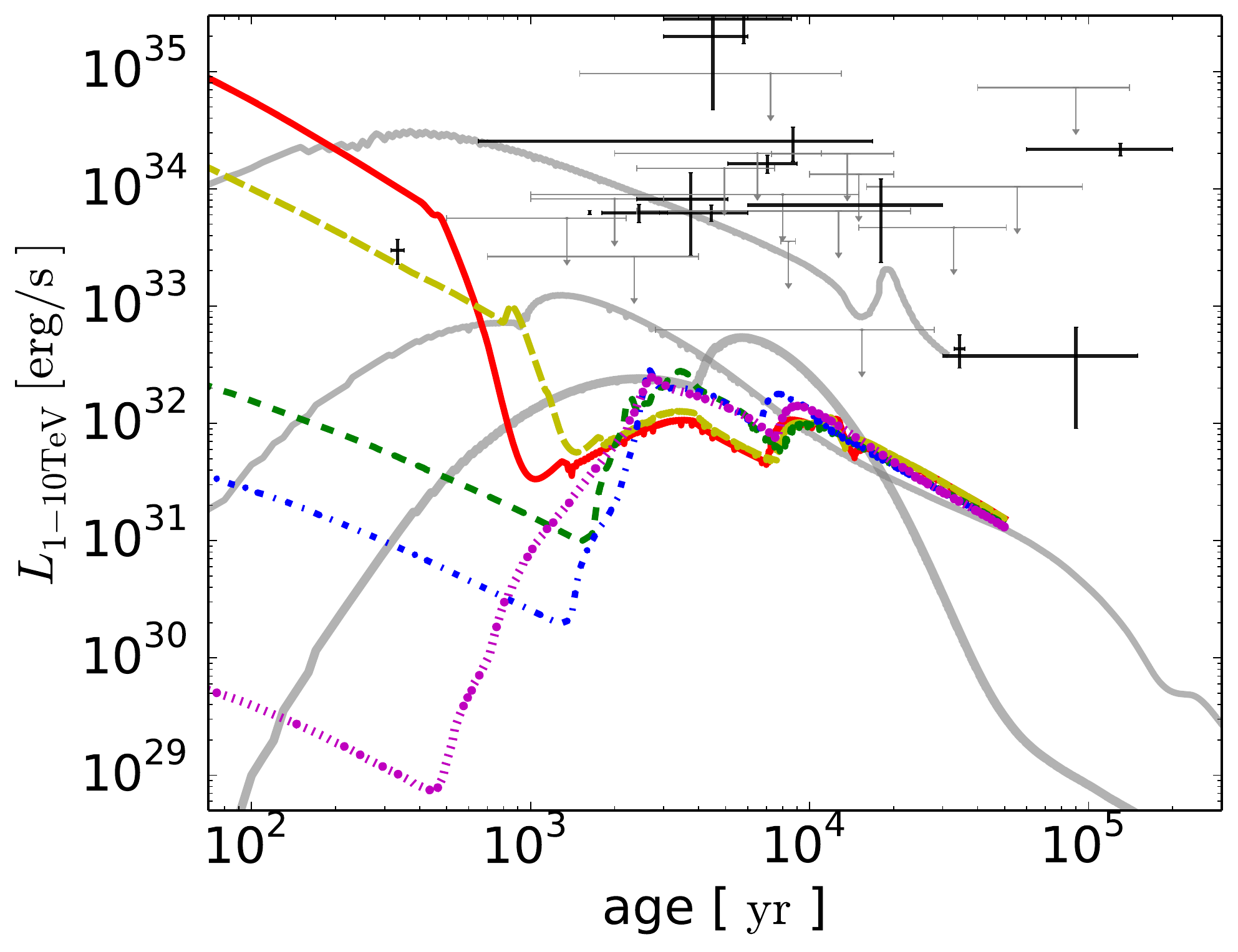}{0.4\textwidth}{(c) 1--10 TeV luminosity}}
    \caption{Same as Figure~\ref{fig:a-lc} but for models in Group B overlaid with CC SNRs observation data. The line formats are the same as in Figure~\ref{fig:b-rv}.}
    \label{fig:b-lc}
\end{figure}
 
In this Section, we switch focus to Gruop B which involves wind-blown structures in the CSM for a massive star progenitor. We compare the results in five models (B1 to B5) in which different mass-loss rates prior to SN are assumed. Figure~\ref{fig:b-rv} shows the hydrodynamic information from the models in analogy to Figure~\ref{fig:a-rv}. We have also overlaid the results from models in Group A for reference. 

The first thing one can immediately notice is that the results are differing from each other mainly in the first $10^3$ yrs or so, after which all models share a very similar dynamical evolution trend. The reason behind this behavior can be explained as follows. Initially, the ejecta expands into the wind structure as shown in Figure~\ref{fig:d-wi} whose typical densities differ according to the mass-loss rates assumed, which explains the difference among the models at young ages before a few $10^2$ yrs old. The initial expansion is also found to be slower in general compared to model A2 which has the same density for the outer uniform medium, due to the much higher gas density immediately outside the ejecta than $0.1$~cm$^{-3}$. As the shock propagates inside the wind with a $r^{-2}$ density profile, the shock decelerates at a slower pace compared to model A2 and the SNRs expands rapidly. For models B1 and B2, the shock experiences a ``break-out'' from a dense wind region into the outer ambient ISM, resulting in an abrupt acceleration of the shock. In the other models, the shock sweeps past the wind bubble until it hits a dense wind shell at a radius determined by pressure balance, and experiences a deceleration until it breaks out from the shell into the outer uniform ISM. In either case (B1--B5) after the ``break-out'', the total swept-up mass behind the shock eventually becomes larger than the ejecta mass which is $12.2~$M$_\odot$ for our progenitor model, the SNR begins to enter the Sedov phase and the dynamics converges among the models regardless of the different mass-loss history and inner CSM structure. We can see that the later evolution of the models is also similar to that predicted by model A2 for the same reason. We note however that there is a possibility that the late-time dynamical evolution can also be highly affected by the mass-loss history of the progenitor if the ejecta mass is smaller than the total moss-loss. For example, the various types of stripped envelope SNe can experience an enhanced mass-loss from binary interactions and so on. In such cases, the SNR should enter the Sedov phase inside the wind instead, which can result in a very different dynamical behavior even after the shock has propagated into the outer ISM region. 

Likewise for Group B, we have overlaid the observation data from a selection of CC SNRs (from Table~\ref{tab:obs}) on the plots, and see a broad range of SNR radii and shock velocities from the population. This is not a surprise because we expect a rich diversity in the CC progenitor types and their associated mass-loss histories and hence CSM environments which cannot be encompassed by the parameter space in our models here. Many are also known to be interacting with giant molecular clouds, especially for the middle-aged remnants. With this in mind, despite the existence of a few outliers, the observed evolutionary trend is not inconsistent with our model predictions. 



Figure~\ref{fig:b-sed} shows the broadband SED for models B1, B3 and B5. Following the convention in \citet{2021ApJ...919L..16Y,Yasuda_2022}, we plot the SEDs at four ages corresponding to the different phases of environment encountered by the forward shock in each model, i.e., the ``$r^{-2}$ wind phase'', the beginning of ``wind-shell interaction phase'', the ``ISM phase'', and at the end of the simulation. Interestingly, at $5 \times 10^4$~yrs old the SEDs are similar to each other and to model A2 for the same reason explained above for the dynamical evolution. 
A slight difference with model A2 exists which is due to the different assumed energetics in the ejecta. 

Differences among the models are mainly observed in the wind phase and shortly after the shock has broken into the outer ISM region. The hadronic versus leptonic origin of the gamma-rays is in line with the mass-loss rates assumed in the models. For model B1 (and B2 not shown here), we can see a double-bump feature in the synchrotron and IC components in the ISM phase. As mentioned above, the shocks in these models experience a ``break-out'' from a dense wind region as it enters the outer ISM. The sudden expansion of the SNR and acceleration of the shock result in a boost in the maximum momenta of the freshly accelerated CRs at the shock in the ISM, and a fast adiabatic energy loss for the CRs accelerated in the past from the shocked wind \citep[e.g.,][]{1984MNRAS.208..645I,1989MNRAS.236..885I}. 
Meanwhile, the smaller magnetic field strength in the ISM compared to that in the dense wind weakens the effect of synchrotron loss on the electrons. Overall, these lead to the appearance of a small bump in the SED at the higher photon energies, which can also be seen in \citet{2021ApJ...919L..16Y,Yasuda_2022,2022ApJ...926..140S}. The difference in the normalization between the bumps comes from the difference in the masses in the shocked wind ($8$~M$_\odot$) and the shocked ISM at the moment.

Figure~\ref{fig:b-lc}(a) shows the light curves at 1 GHz from Group B. 
The luminosities in all models decrease from the beginning in the ``$r^{-2}$ wind phase'' in contrast with Group A, and they decrease in a similar fashion among the models in accordance with the decreasing $B$-field and gas density with the shock radius (i.e., $L \propto \rho_0 B_0^2 \propto \rho_0^2$ at the shock where $\rho_0\propto \dot{M}r^{-2}$ with the same wind velocity), until the shock breaks out from either the inner dense wind or the dense shell outside the wind bubble at $\sim 10^3$ yrs. After that, all the models converge to a radio luminosity similar to that predicted in model A2. 
Not surprisingly, the radio light curves after the SNR has entered the ISM phase retain no information from the mass-loss histories of the progenitor (see discussions above for possible exceptions). The ages at which the transition commences differ for each model (400, 2000, 1500, 1100 and 300 yrs old for models B1 to B5 respectively) in accordance with the CSM structure shown in Figure~\ref{fig:d-wi}. We note that since we do not consider the mass-loss history of the progenitor derived from stellar evolution models in this parametric study, these ages can alter when a more self-consistent stellar evolution model is taken into account. 

In the $r^{-2}$ wind phase, the gamma-ray luminosity decreases with time as shown by the light curves in Figure~\ref{fig:b-lc}(b) and \ref{fig:b-lc}(c). During this phase, the gamma-ray emission is mostly dominated by the $\pi^0$-decay component as shown in Figure~\ref{fig:b-sed}, whose flux ($\propto \rho_0^2$) decreases monotonically with age. This is consistent with the results presented in \citetalias{2019ApJ...876...27Y} \footnote{The $\pi^0$-decay component tends to be more luminous in our models than \citetalias{2019ApJ...876...27Y} due to the fact that our ejecta mass ($12.2$~M$_\odot$) is larger than that used in \citetalias{2019ApJ...876...27Y} ($3$~M$_\odot$) but with a similar explosion energy. The resulted slower expansion of the SNR leads to a higher ambient wind density encountered by the shock at a given age.}. For models with a small mass-loss rate such as model B5, the IC component whose flux has a weaker dependence on the wind density ($\propto \rho_0$) can become comparable to the hadronic contribution near the end of this phase, leading to a more gradual decay of the gamma-ray luminosity for such models. The major differences with \citetalias{2019ApJ...876...27Y} begin to appear as the shock leaves the freely expanding wind and enters the wind shell and outer ISM, which were not considered in \citetalias{2019ApJ...876...27Y}. Likewise with the radio counterpart, the gamma-ray light curves merge into one similar to that of model A2. 

Compared with our models, especially at older ages, the observed CC SNRs (Table~\ref{tab:obs}) show relatively high radio and gamma-ray luminosities, probably due to a higher average ISM density encountered by the remnants (e.g., molecular clouds) and hence also higher magnetic fields than Group B. In fact, the observation data mostly fall between or above the results from models A1 and A2, suggesting a higher ambient gas density than the typical warm ISM phase. 
Our models suggest that observations of CC SNRs at younger ages are the most effective in probing the surrounding CSM environment. However, 
recent hydrodynamic simulations with stellar evolution models and the associated CSM structures taken into account self-consistently \citep[e.g.,][]{2019ApJ...885...41M, 2021ApJ...919L..16Y, Yasuda_2022} have shown that the non-thermal emission of a SNR at different evolutionary stages heavily depends on the progenitor type and its pre-SN activities, suggesting a promising prospect of future observations of non-thermal emission on constraining the progenitor origin of SNRs (see Section~\ref{subsec:res-force} below for an example based on our models). 
The incorporation of different progenitor types and their association with surrounding environments expected for CC SNRs will be accounted for based on stellar evolution models in a followup paper. 

\subsection{Effects in the radiative phase}\label{subsec:res-rad}

Here we present results showing the impact of physical processes in the radiative phase on the calculated emission spectra. We will mainly focus on models in Group A for illustration purpose. 

\subsubsection{Re-acceleration effect}\label{subsec:res-reacc}

\begin{figure}[ht]
    \gridline{\fig{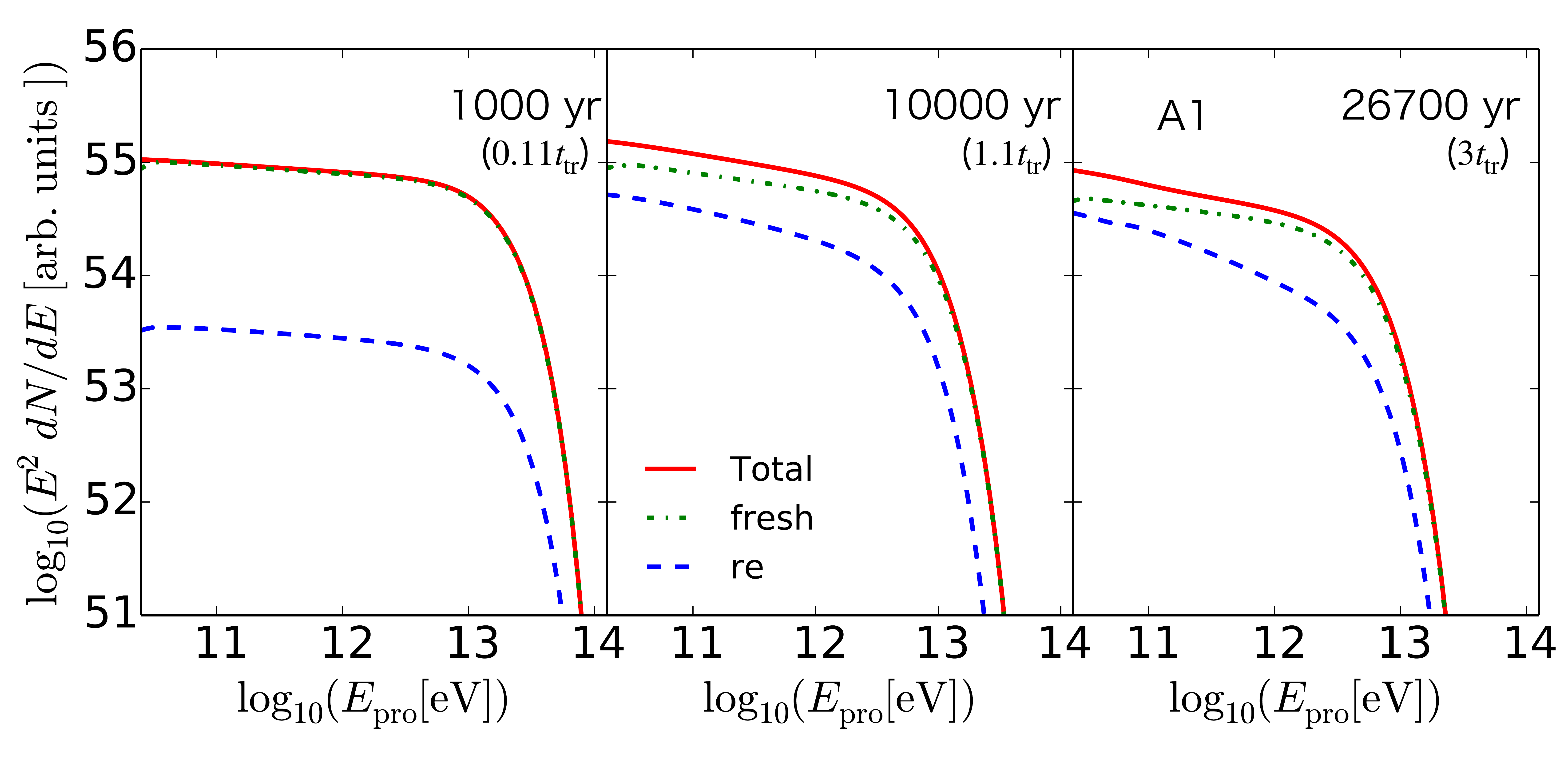}{0.5\textwidth}{(a) Case of A1}}
    \gridline{\fig{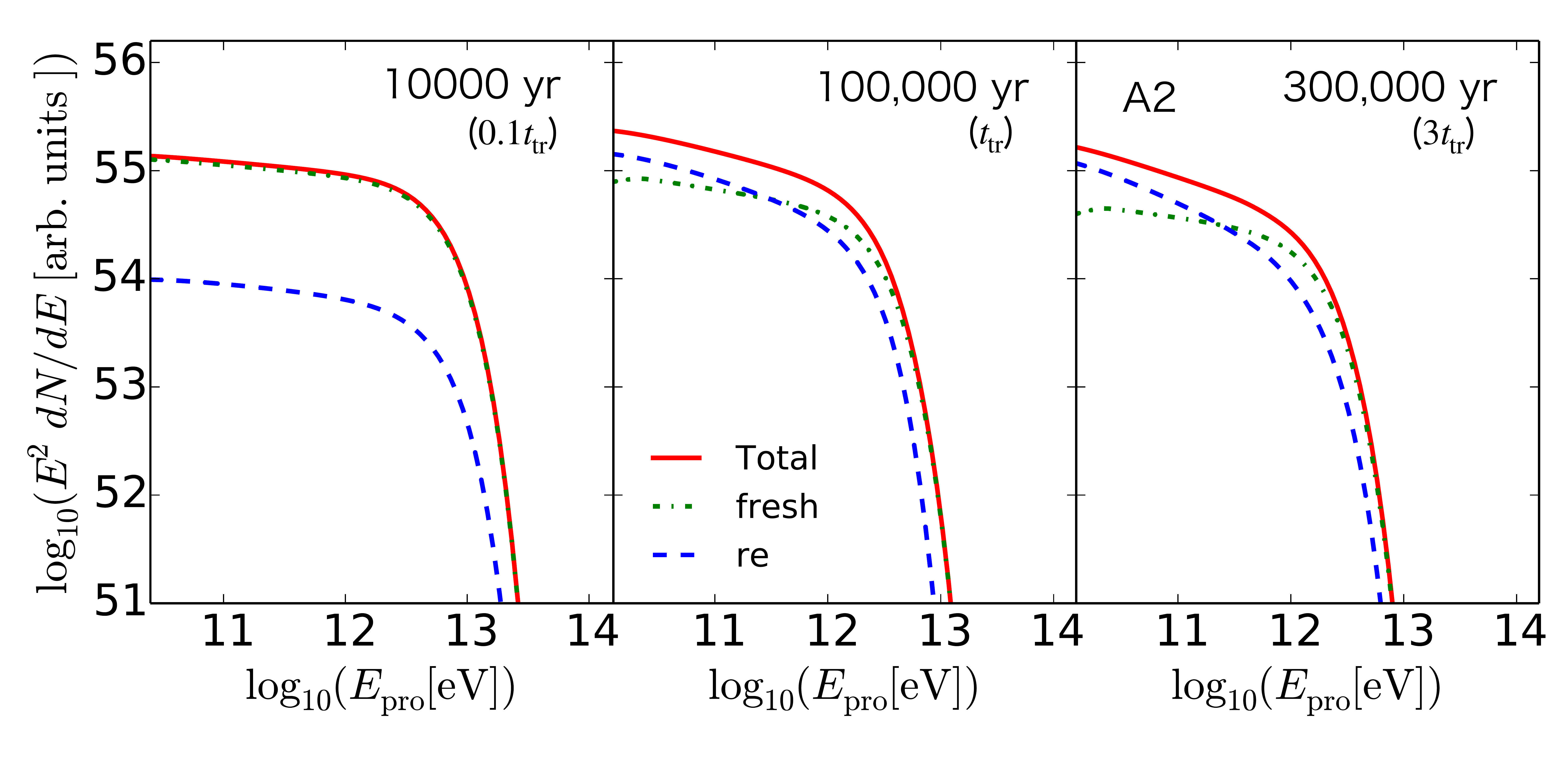}{0.5\textwidth}{(b) Case of A2}}
    \caption{Time snapshots of CR proton spectra at three different dynamical ages integrated over the whole SNR volume for models A1 and A2. The total spectra (red solid) is decomposed into components from the freshly accelerated CRs (green dashed-dotted) and the re-accelerated CRs (blue dashed-dotted) to visualize their relative contributions to the accelerated CRs. 
    }
    \label{fig:a-fpeDSAvol}
\end{figure}

\begin{figure}[ht]
    \plotone{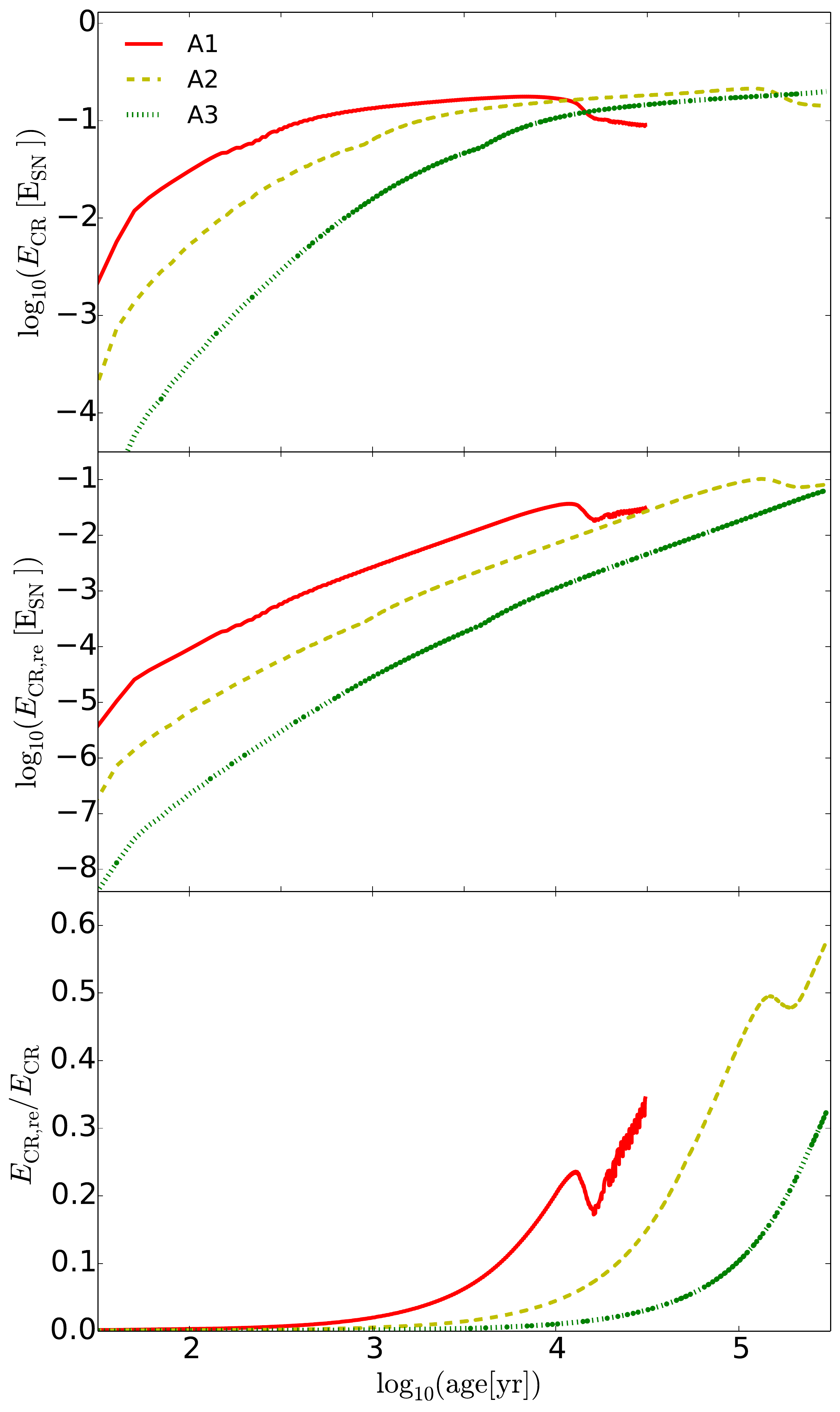}
    \caption{The total kinetic energy in the accelerated CRs integrated over the SNR volume as a function of age for models A1 to A3. The line formats are the same as in Figure~\ref{fig:a-rv}. The upper panel shows the time evolution of the total energy in all CRs inside the SNR in units of the SN explosion energy $E_\mathrm{SN}$. Likewise the middle panel shows the evolution of the re-acceleration component. The lower panel shows the energy ratio between the re-acceleration component and the total.}
        \label{fig:a-ecr}
\end{figure}

\begin{figure}[ht]
    \plotone{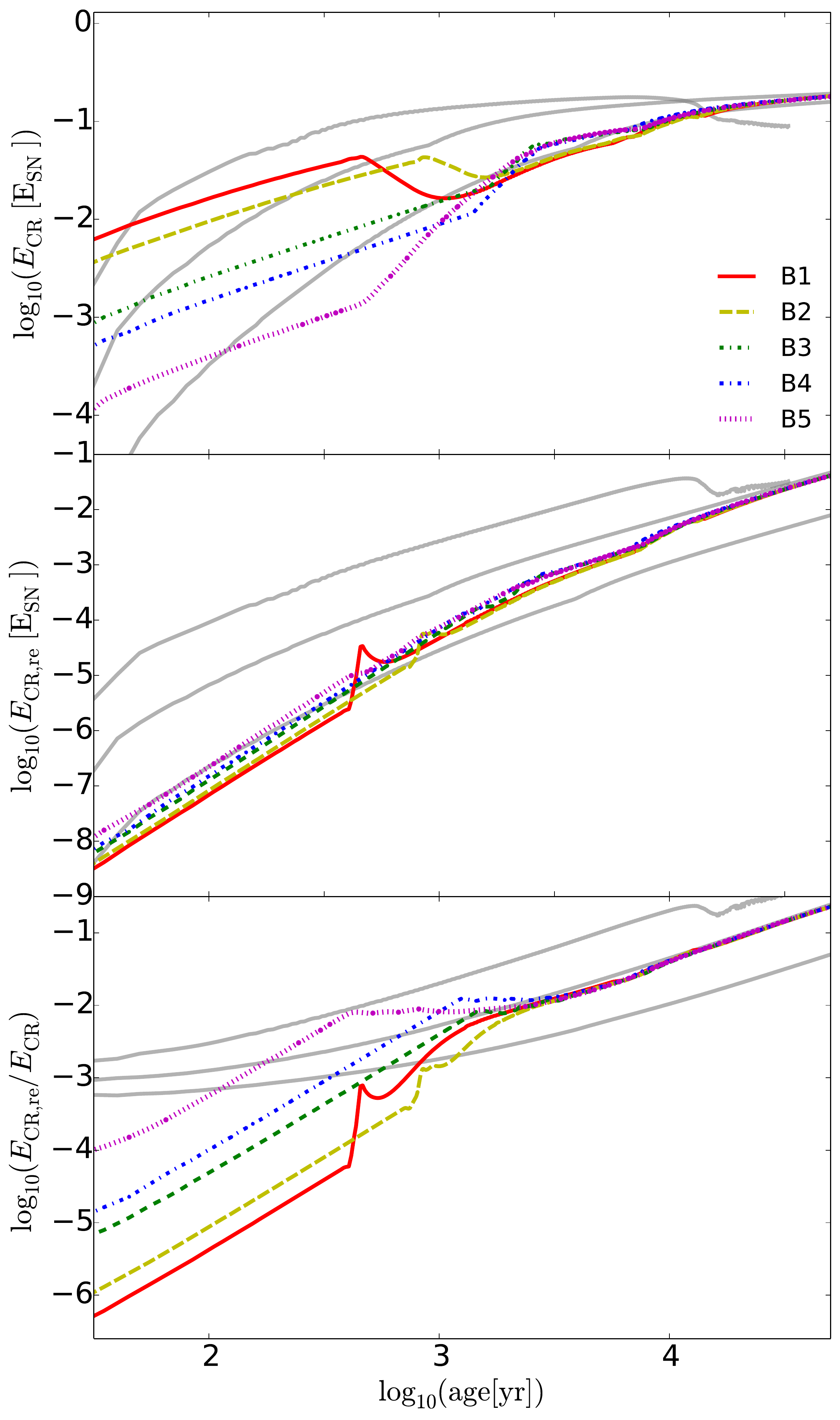}
    \caption{Same as Figure~\ref{fig:a-ecr} but for the models in Group B. The line formats are the same as in Figure~\ref{fig:b-rv}. }
        \label{fig:b-ecr}
\end{figure}

As the SNR shock decelerates and eventually becomes radiative, the ability of the shock in injecting and accelerating particles from thermal energies becomes weak. On the other hand, it has been shown that re-acceleration of pre-existing relativistic particles such as the Galactic CRs can remain effective at fast radiative shocks, which can take over as the dominant mechanism of non-thermal emission in more evolved SNRs. This effect from re-acceleration of pre-existing CRs in older SNRs with shock-cloud interactions have been investigated by a few recent works \citep[e.g.,][and reference therein]{2010ApJ...723L.122U, 2015ApJ...806...71L,galaxies7020049}, which suggest that the bright GeV gamma-ray and radio emission observed in middle-aged SNRs such as W44 and IC 443 can be mostly accounted for by the re-acceleration of Galactic CRs at their fast radiative cloud shock\footnote{This scenario has been questioned along the line of the estimated pre-shock and post-shock cloud masses being unreasonably large for remnants such as G39.2-0.3 and W44 \citep[see, e.g.,][]{2020MNRAS.497.3581D}. However, the shocked cloud mass usually cannot be estimated trivially since in such a scenario most of the non-thermal emission will be coming from a very spatially confined dense radiative cold shell behind the FS instead of a considerable fraction of the post-shock volume. This usually leads to an over-estimation of the volume filling factors and so on in simple order estimations, and hence unreasonably high masses associated with the gamma-ray luminosities. This has been explained in the beginning of Section 5 in \citet{2016AA...595A..58C}, and the estimated total mass of the shocked gas responsible for the gamma-ray emission in W44 is much smaller in \citet{2015ApJ...806...71L} than the values cited in \citet{2020MNRAS.497.3581D} for example.}. These studies, however, initialized the cloud shock in an ad-hoc manner without considering the dynamics of the ejecta from explosion as well as in the earlier evolutionary stages before the SNR hits a dense cloud. This may lead to a failure in capturing the effects from the long-term evolution of the SNR and the complete history of CR acceleration from explosion to the current day. By including the essential physics similar to these previous works, our long-term simulations can serve to remedy this problem.



Based on \citet{2010ApJ...723L.122U}, we take into account the re-acceleration of pre-existing cosmic rays in parallel to the injection of thermal particles in the DSA process throughout the whole lifetime of a SNR until its shock dies out. 
In addition, by adding this effect to our self-consistent calculations, we can more accurately estimate the total amount of CRs accelerated through the lifetime of a SNR, thus helping us evaluate quantitatively the contribution of re-acceleration to the total CR output from a remnant as a function of age. 

To show the fractional contribution of re-acceleration to the accelerated CRs, Figure~\ref{fig:a-fpeDSAvol} displays snapshots of the accelerated proton spectra integrated over the whole SNR volume in certain selected ages. For a quantitative discussion, we also adopt the time evolution of the total kinetic energy from each CR component inside the SNR\footnote{The escaped CRs are not included here since they do not contribute to the non-thermal emission from the remnant.}, i.e., $E_\mathrm{CR,re}$ for the kinetic energy in the re-acceleration component and $E_\mathrm{CR,fresh}$ in the freshly accelerated CR component respectively, which is shown in Figure~\ref{fig:a-ecr}. We calculate $E_\mathrm{CR,i}$ (where $\mathrm{i}=\{\mathrm{re},\ \mathrm{fresh}\}$) as 
\begin{eqnarray}
&E&_\mathrm{CR,i}=\int\int (\gamma-1)m_\mathrm{p} c^2 f_\mathrm{i}(x,p) 4\pi p^2dp d^3x\\
&E&_\mathrm{CR}=E_\mathrm{CR,re}+E_\mathrm{CR,fresh},
\end{eqnarray}
where $\gamma$ is the Lorentz factor, $f(x,p)$ is the phase-space distribution function of the non-thermal particles. 
From Figure~\ref{fig:a-fpeDSAvol} and the top and middle panels of Figure~\ref{fig:a-ecr}, the flux of the re-acceleration component remains approximately constant from 1 $t_\mathrm{tr}$ to 3 $t_\mathrm{tr}$, whereas the flux of the freshly accelerated component decreases as the shock weakens and the injection and acceleration becomes inefficient, and is now dominated by the CRs accelerated in the past which is suffering from adiabatic loss. We note that at a certain age, the fresh and re-accelerated CR populations are both composed of the accumulation of particles with spectra of different indices as they were accelerated by the shock at different velocities at different ages. Furthermore since the time evolution of the acceleration efficiency of the fresh versus re-accelerated CR are typically different in our models, the resulting overall spectral index can be different as well. 

As shown in the bottom panel of Figure~\ref{fig:a-ecr}, the ratio between the re-acceleration component and the total CR content increases with the SNR age up to a few $t_\mathrm{tr}$, indeed indicating an increasing importance of re-acceleration effect in older objects. 
However, the ratio increases only up to $\sim 35\%$ (A1 and A3) and $\sim 60\%$ (A2) by $3t_\mathrm{tr}$, 
which is far from a total domination used by the previous works. 
Admittedly these numbers should depend on parameters such as the ISM density, mass-loss history and so on as is shown by the differences between models A1 to A3, but our results clearly illustrate that it is important to account for the CR acceleration history coherently starting from the explosion itself in order to obtain an accurate estimation of the CR energy budget and spectra, and hence the non-thermal emission properties.


For a quick comparison, we also show the results from the models in Group B in Figure~\ref{fig:b-ecr}, and found that the final ratio is close to that predicted by model A2 (with the same $n_\mathrm{ISM}=0.1\ \mathrm{cm}^{-3}$). 
In the young wind-interaction phase, the total (and fresh) component is roughly proportional to the pre-SN mass-loss rates, which can be understood as coming from the different masses in the gas swept up by the shock (and hence the number of particles injected into DSA) at any given age. On the other hand, $E_\mathrm{CR,re}$ shows a much weaker dependence on the wind properties mainly due to the different nature of the seed particles involved, i.e., the pre-existing Galactic CRs whose density profile is assumed to be not affected by the mass loss. We can still observe a slight difference among the models which scales inversely with the mass-loss rate, and can be interpreted as coming from the small difference in the shock velocities. These different behaviors between $E_\mathrm{CR,re}$ and $E_\mathrm{CR,fresh}$ lead to an interesting outcome that the ratio $E_\mathrm{CR,re}/E_\mathrm{CR}$, as shown in the bottom panel in Figure~\ref{fig:a-ecr} and \ref{fig:b-ecr}, scales with the upstream gas density in an almost opposite way from the total $E_\mathrm{CR}$ in the young phase. 
When the shock is still strong, propagating in the stellar wind, $E_\mathrm{CR}$ is mostly dominated by the freshly accelerated particles. This is quite different from the situation we expected for older SNRs which are often found to be interacting with dense molecular clouds and the shocks have already become radiative. In the latter case, $E_\mathrm{CR,re}$ is expected to play a much more important role than in the younger phase. At around a few $10^2$ to $10^3$~yrs, however, the ratio can reach around $1\%$ for model B5, implying that re-acceleration of pre-existing CRs does contribute to the non-thermal emission for progenitors with smaller mass-loss rates. 
Here we have ignored the possibility of the evacuation of low-energy Galactic CRs by the magnetized stellar wind before the SN explosion and hence a reduced injection of pre-existing CRs in the wind region. However, as one can see in the leftmost panels of Figure~\ref{fig:a-fpeDSAvol}(a,b), the contribution of the re-accelerated CRs is small compared to the total component (lower than 10\%), which means that re-acceleration during the younger phase does not affect the overall broadband emission in a significant way.

As noted in the beginning of this Section, while it has been believed that the re-acceleration of Galactic CRs is sufficient to explain the non-thermal emission in older SNRs interacting with high density environments,
our long-term simulation indicates that any CRs accelerated in the past before shock-cloud interaction cannot be ignored and should be treated self-consistently with the hydrodynamic evolution of the remnant from the explosion up to the current epoch, even though the shock has already become radiative now. 

\subsubsection{Spectral break due to ion-neutral damping}\label{subsec:res-neutral}

\begin{figure}[ht]
    \gridline{\fig{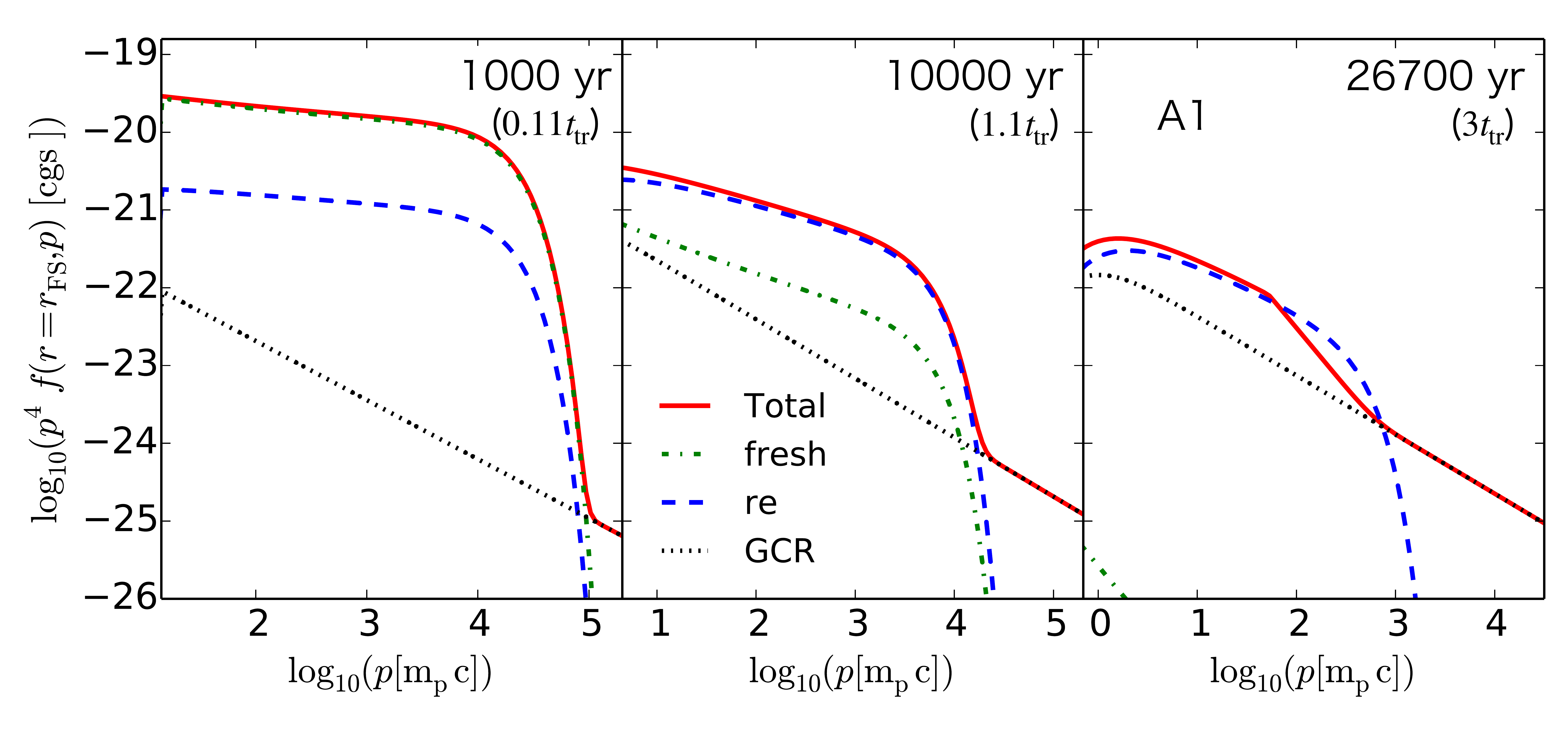}{0.5\textwidth}{(a) Case of A1}}
    \gridline{\fig{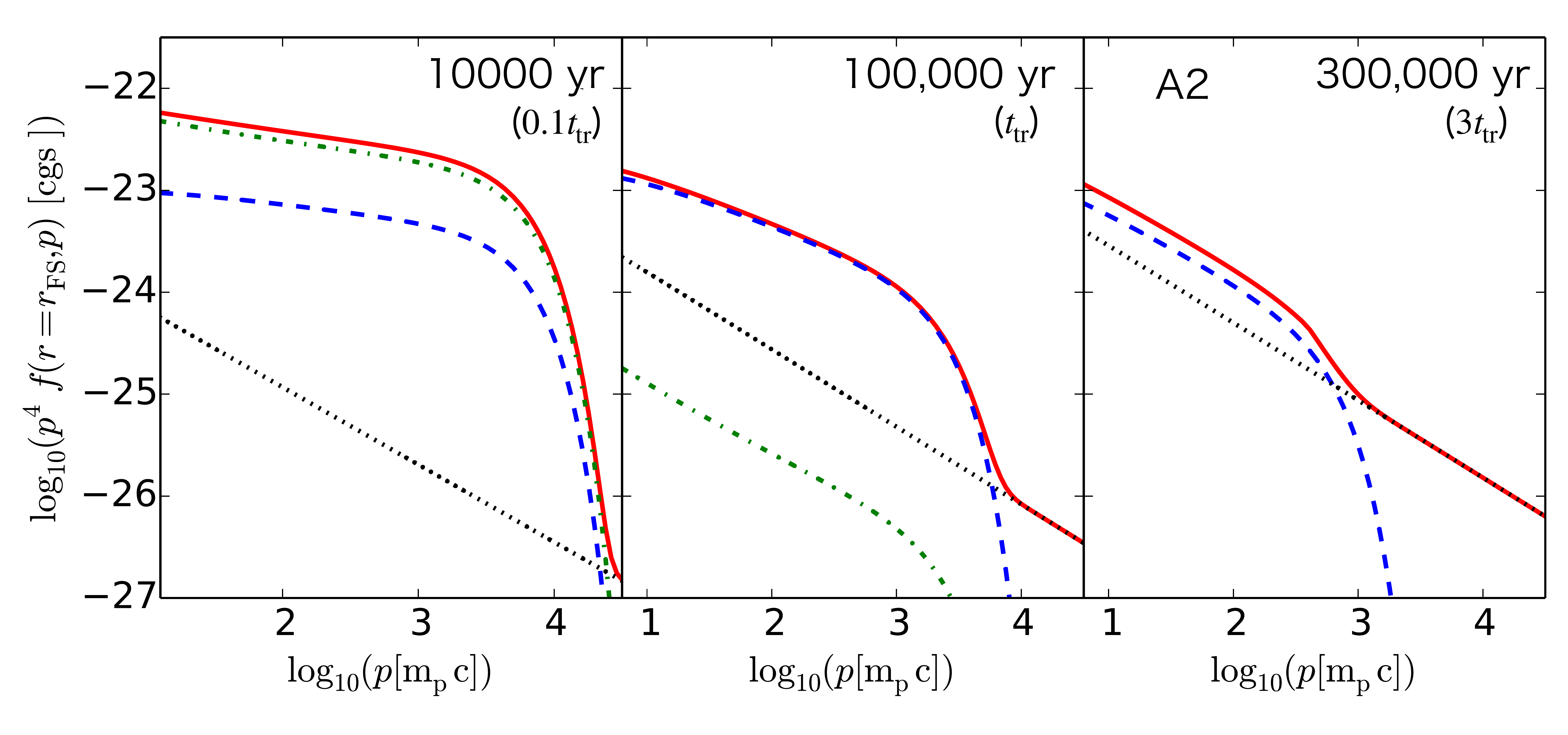}{0.5\textwidth}{(b) Case of A2}}
    \caption{Local phase space distribution of CR protons at the shock plotted at three time epochs. Same as Figure~\ref{fig:a-fpeDSAvol}, the total spectrum (red solid) is decomposed into components from the freshly accelerated CRs (green dashed-dotted) and the re-accelerated CRs (blue dashed-dotted). The spectral break from ion-neutral damping effect is applied to the total spectrum only for clarity. The background Galactic CRs is also plotted for reference (grey dotted line). }
        \label{fig:a-fpeDSA}
\end{figure}
 
  \begin{figure*}[ht]
	\centering	
	\plotone{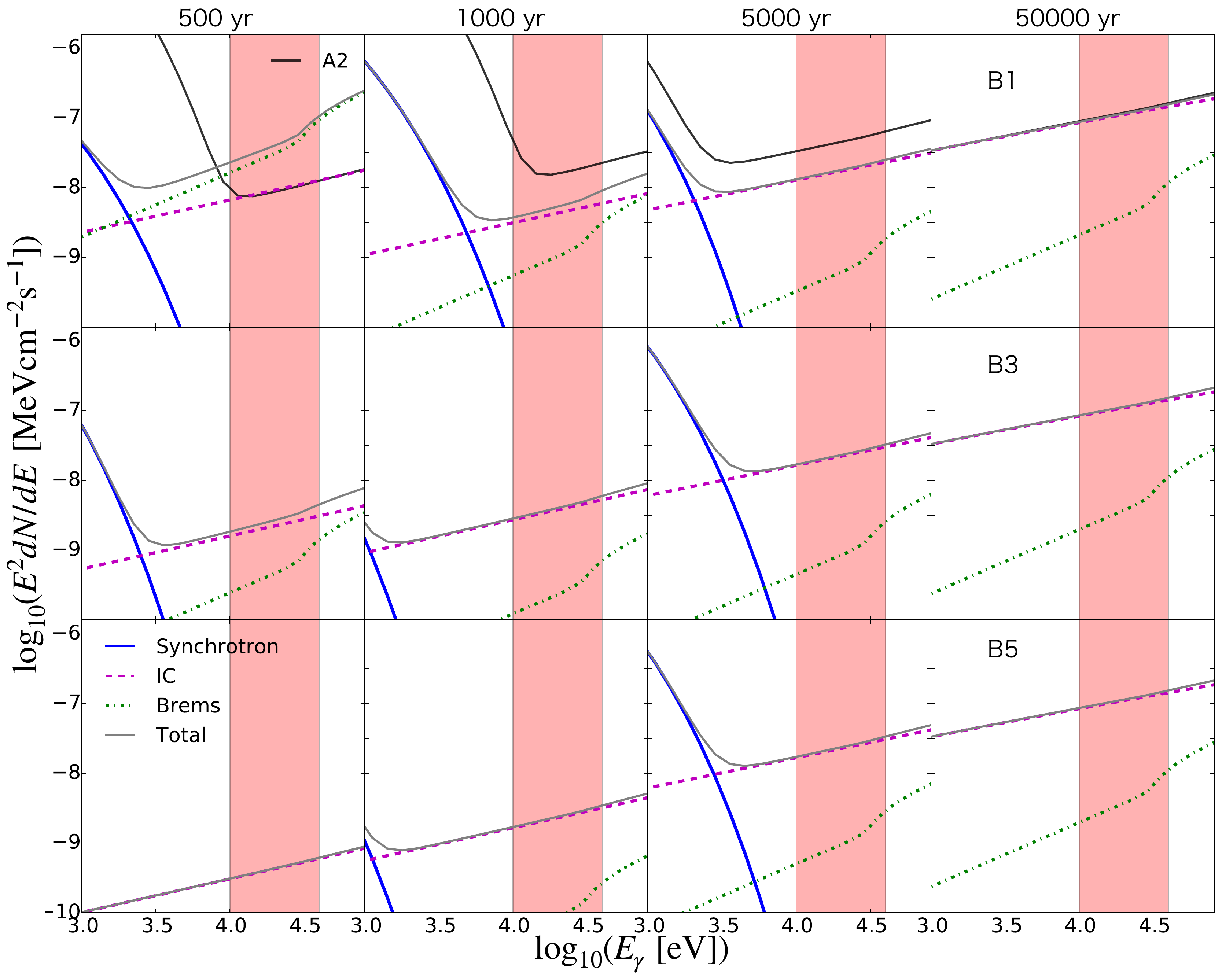}
    \caption{Same as Figure~\ref{fig:b-sed} zoomed into the energy range of $1-80\ \mathrm{keV}$. In the top panel, the total SED of Model A2 is overlaid as a black solid line. The red shaded area indicates the $10-40\ \mathrm{keV}$ range in which future observations by \textit{FORCE} is optimal.}
	\label{fig:b-sedfor}
 \end{figure*}

As mentioned above, we cannot confirm any clear break feature caused by ion-neutral damping effect in the volume-integrated spectra in Figure~\ref{fig:a-fpeDSAvol}. But this does not necessarily mean that ion-neutral damping is not happening at all, and in some models in this study, ion-neutral damping indeed takes effect. 

Figure~\ref{fig:a-fpeDSA} shows the local proton spectra accelerated at the shock (i.e., without the CRs accelerated in the past in the downstream) separated into  the freshly accelerated and re-accelerated CR components. Panel (a) shows the result of model A1 up to an age of 26,700 yr ($3t_\mathrm{tr}$), where we can see that the re-accelerated CRs becomes the dominant component after 10,000 yrs ($1.1t_\mathrm{tr}$),
and model A2 in panel (b) shows a similar behavior at the same dynamical ages. 
At $t = 3t_\mathrm{tr}$, we can indeed see a spectral break feature at the momentum $p_\mathrm{br}\lesssim10^2\ \mathrm{m_p}c$ for model A1 and $p_\mathrm{br}\sim \mathrm{a\ few}\ 10^2\ \mathrm{m_p}c$ for model A2, which comes from the ion-neutral damping effect\footnote{Note that we intentionally apply the spectral break to the total spectra (red lines) only but not to the individual components in order to illustrate the effect of the spectral break on the shape of the particle distribution.}. This feature does not appear in the photon spectra (Figure~\ref{fig:a-sed}) since in the context of our model parameter space, the total CR spectra are mainly dominated by the particles accelerated in the past before the shock has become radiative, consistent with our discussion above on the CR energy budget.

The presence or absence of a spectral break in the gamma-ray spectrum depends on the detailed structure of the ambient environment.
In our study, the maximum ISM density covered by the parameter space is $n_\mathrm{ISM}=10\ \mathrm{cm}^{-3}$ and is assumed to be uniform in space. In many older SNRs, the shocks are currently interacting with molecular clouds with a density at least an order-of-magnitude higher. A more realistic environment may also contain a moderately dense region in the vicinity of the ejecta and denser clouds in the outer region at a few $10~\mathrm{pc}$ \citep{2015SSRv..188..187S}. Assuming such an environment, it is possible that we can see a pronounced ion-neutral break feature in the overall gamma-ray spectra at late times when the emission from the shock-cloud interaction region becomes the dominant component in the SED. We will consider such a situation in a future work.

\subsection{Prospects for \textit{FORCE}}\label{subsec:res-force}

 \begin{figure}[p]
    \gridline{\fig{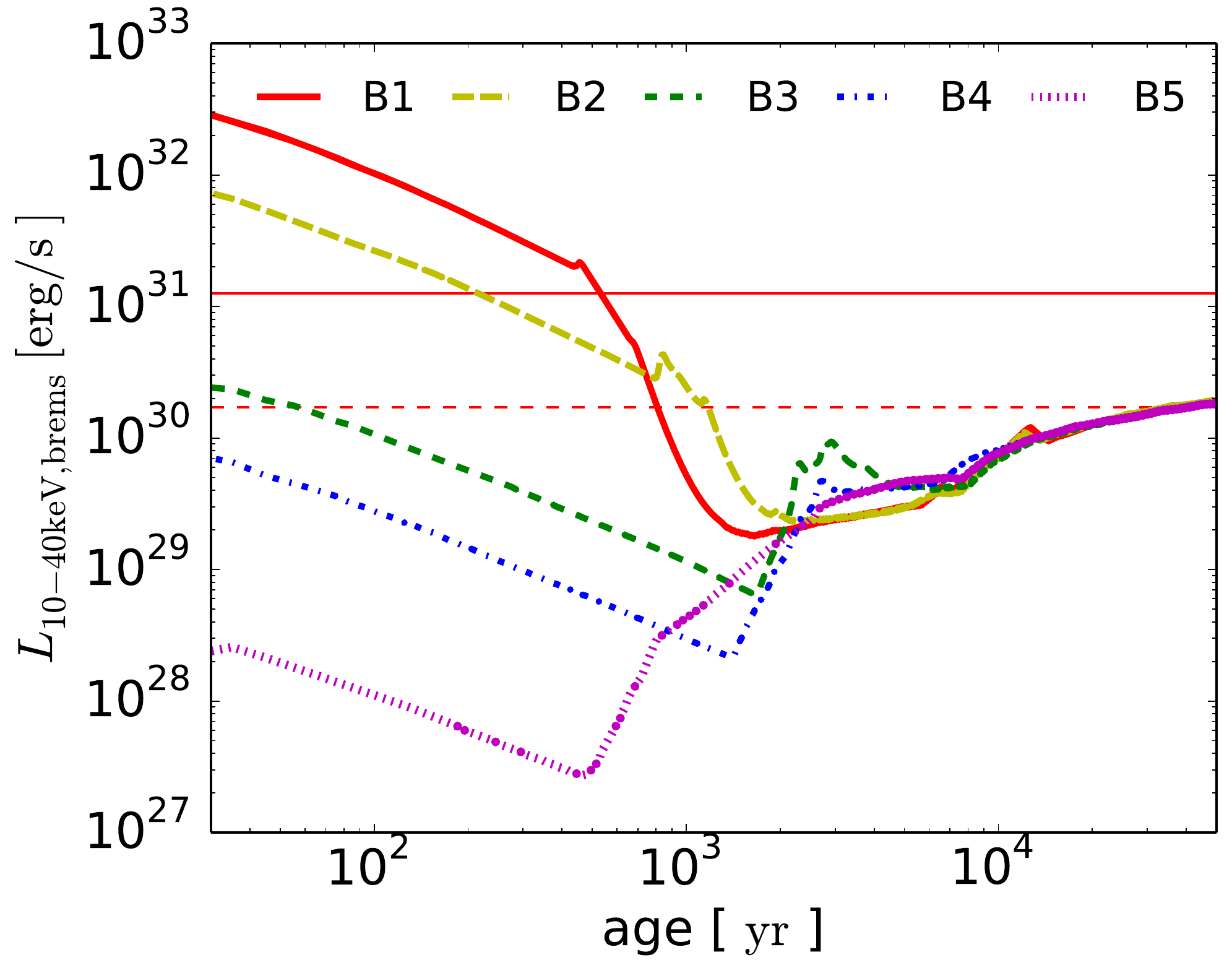}{0.4\textwidth}{(a) Non-thermal bremsstrahlung}}
    \gridline{\fig{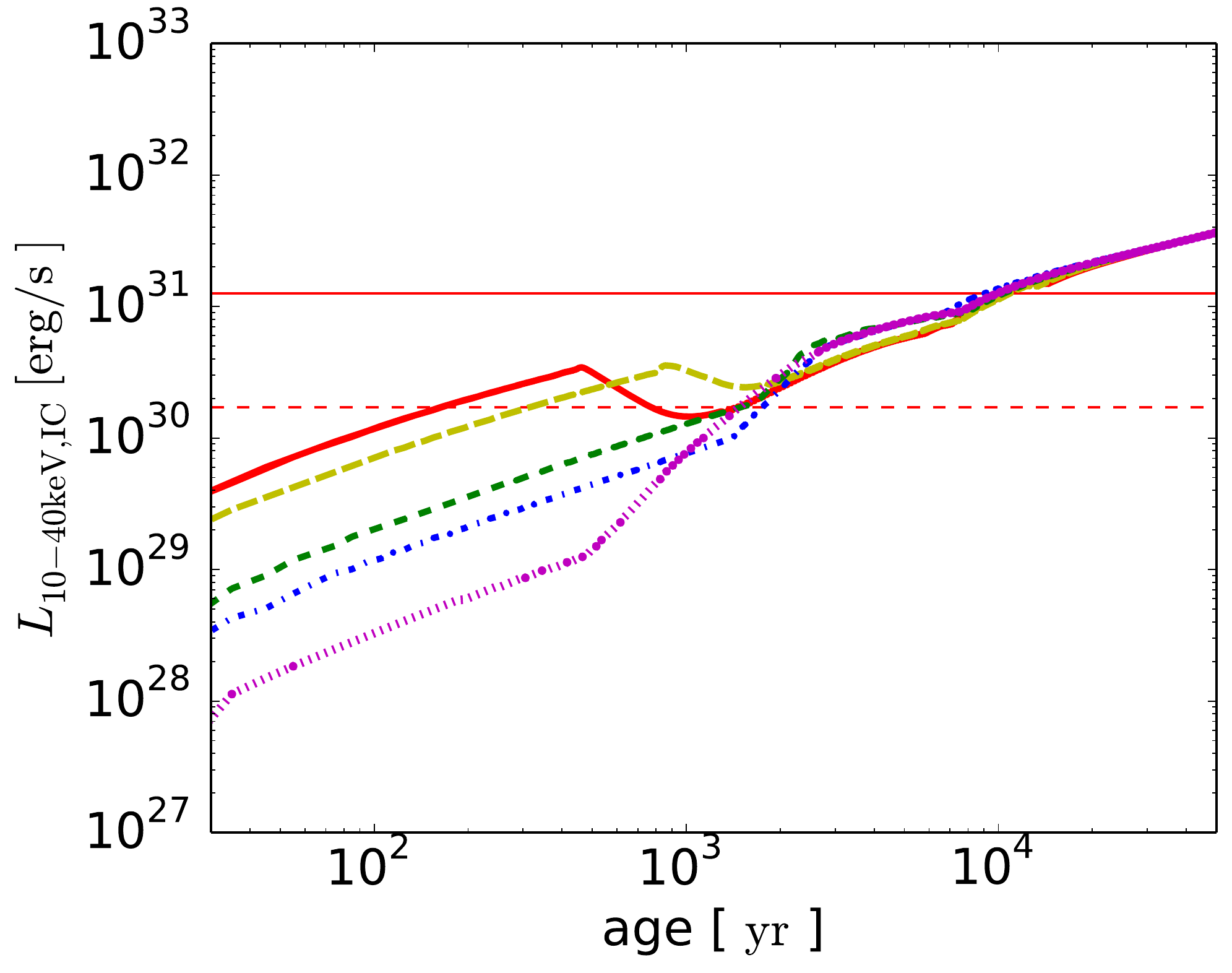}{0.4\textwidth}{(b) Inverse-Compton}}
    \gridline{\fig{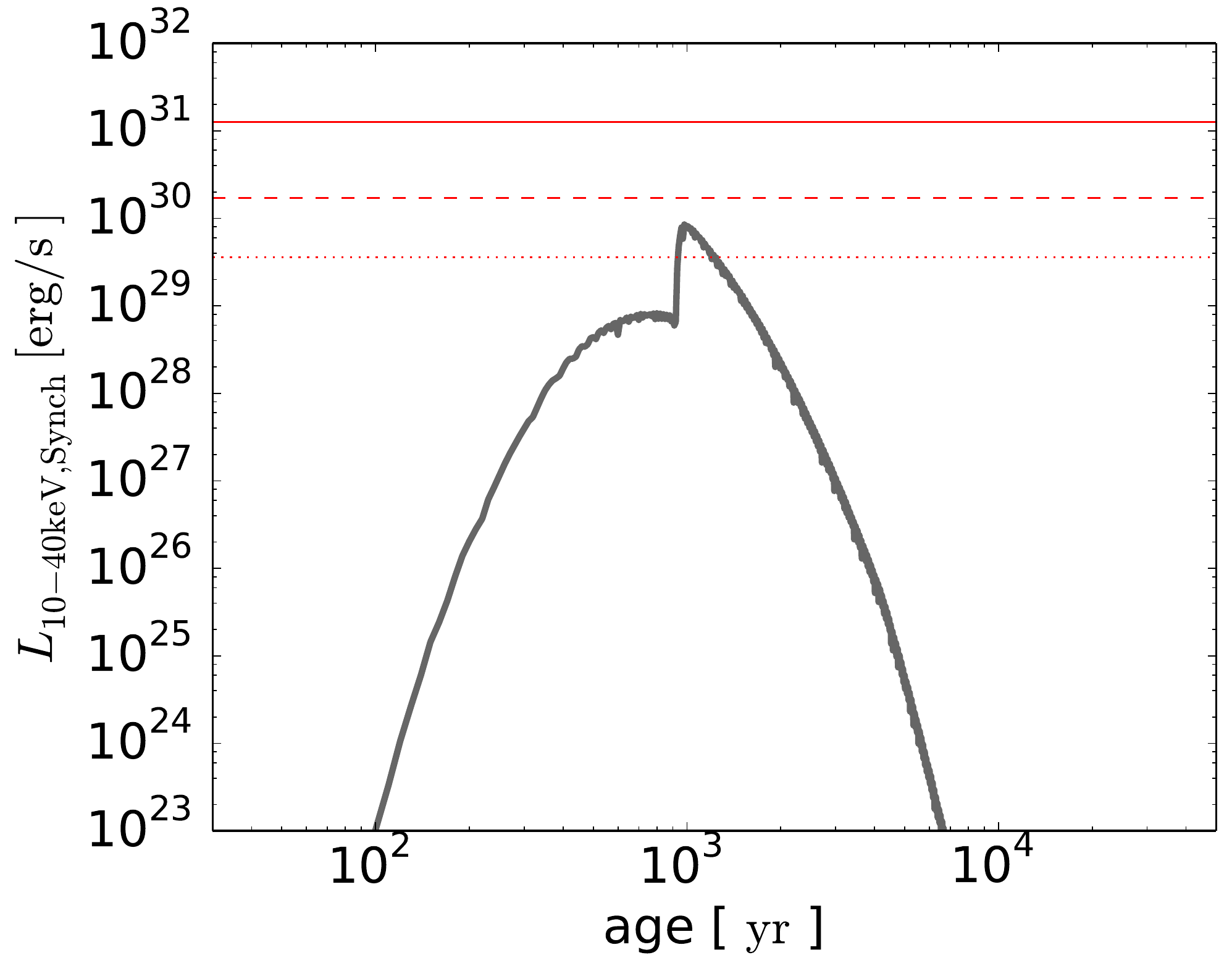}{0.4\textwidth}{(c) Synchrotron (model A2)}}
    \caption{Panel (a) \& (b): Light curves in the $10-40\ \mathrm{keV}$ energy range separated into the non-thermal bremsstrahlung (a) and IC (b) components. The line formats are the same as in Figure~\ref{fig:d-wi}. The horizontal solid, dashed and dotted lines indicate the $3.5\sigma$ sensitivity of \textit{FORCE} at 1~kpc with exposure times of $10^4$, $10^5$ and $10^6$ s, respectively. Panel (c): same but for the synchrotron component from model A2 in the black solid line.}
    \label{fig:b-lcfor}
\end{figure}

Before we end our discussion, we will employ our models to assess the prospects of \textit{FORCE} (\textit{Focusing On Relativistic universe and Cosmic Evolution}) on SNR research, which is a next-generation space hard X-ray imaging and spectroscopy instrument planned to be launched in the near future (later 2020's). This instrument excels at hard X-ray imaging with a superior angular resolution ($<15''$) and possesses a large effective area at energies $> 10\ \mathrm{keV}$ to ensure high photon statistics for spatially resolved spectroscopy (\url{https://www.cc.miyazaki-u.ac.jp/force/wp-content/uploads/force_proposal.pdf}). In this Section, we will compare our results with the sensitivity of \textit{FORCE} in the 10--40 keV band to discuss possible science achievable by this future mission. For a discussion on the \textit{Cherenkov Telescope Array (CTA)} for TeV observations, we refer the readers to \citetalias{2019ApJ...876...27Y}.

Figure~\ref{fig:b-sedfor} shows the SED from model B1, B3 and B5 in the $1-80\ \mathrm{keV}$ X-ray band which will be covered by the baseline design of \textit{FORCE}. The $10-40\ \mathrm{keV}$ band is shown by the shaded region in red. First of all, the synchrotron components in our CC SNR models are found to be faint in this energy band, due to a weak averaged magnetic field inside the wind during the younger stage, and a low energy cutoff from synchrotron loss during the ISM phase. We thus focus the discussion of the non-thermal X-ray observation on the two other components of IC and non-thermal bremsstrahlung. Within the context of our models, the SED evolution is almost homologous from an age of a few $10^3$ yrs (see Figure~\ref{fig:b-lc}), so in order to extract information like mass-loss histories observations of younger remnants are necessary. For reference, the total SED from model A2 is plotted in top panel. The result suggests that young ($\sim$ 1,000 yrs old) Type Ia SNRs interacting with a tenuous environment may emit synchrotron radiation well into the $>10$~keV range. 

Figure~\ref{fig:b-lcfor} shows the model X-ray light curves in the $10-40\ \mathrm{keV}$ band compared with the $3.5\sigma$ sensitivities for a point source with an angular resolution of $15''$ and exposure times of $10$ and $100$ ks (\url{https://www.cc.miyazaki-u.ac.jp/force/wp-content/uploads/force_proposal.pdf}). Here a distance of 1~kpc is assumed. The bremsstrahlung luminosity up to around $10^3\ \mathrm{yrs}$ for the models with a high mass-loss rate such as B1 and B2 here are bright compared to the sensitivity curve. A future detection of this component will bring about information on the CSM structure and mass-loss histories of the progenitors. Moreover, the non-thermal bremsstrahlung depends on the upstream gas density $\rho_\mathrm{0,H}$ as $K_\mathrm{ep}\rho_\mathrm{0,H}^2$. In the case that one can simultaneously detect the $\pi^0$-decay gamma-ray emission and obtain its flux ratio with the bremsstrahlung component, it is possible to obtain a stringent constraint on the $K_\mathrm{ep}$ parameter to understand the electron injection process in DSA at strong collisionless shocks. 
The IC component becomes bright enough to be detectable at an age older than $\sim10^3\ \mathrm{yr}$ at 1~kpc, during which the emission is dominated by the electrons accelerated in the shocked ISM. 

In contrast with the radio and gamma-ray light curves which decline with age at late time, it is interesting to observe that the hard-X ray luminosities from both the bremsstrahlung and IC components generally rise with age in the ISM phase. This different behavior can be explained by an accumulation of electrons at low energies due to the adiabatic and synchrotron loss of the higher energy electrons, as well as the steepening of the electron spectrum at the shock which weakens as the SNR evolves. This implies that older SNRs, especially those interacting with a dense environment, are good targets for \textit{FORCE}.    
On the other hand, even with a long observation time of 1 Ms, the synchrotron emission is barely detectable by \textit{FORCE} (see panel (c); note that only model A2 is shown here with the highest synchrotron luminosity among the models) although there is a possibility that the synchrotron from secondary electrons/positrons may increase the luminosity to some extent, which is out of the scope of this paper. The synchrotron luminosity drops rapidly with age after the peak at around $10^3$ yrs due to severe synchrotron loss of the electrons close to their maximum energy. The sudden enhancement near the peak comes from the non-linear DSA effect and CR-induced $B$-field amplification as explained above. Despite the difficulty, however, an upperlimit from \textit{FORCE} combined with observations at softer X-rays and other wave bands will serve to constrain spectral models further to obtain possible range of key parameters such as the magnetic field strength and maximum electron energy. 

With a high spatial resolution, non-thermal SNRs with angular sizes of  $\sim 10$ arcmin or bigger such as the Galactic SNR RX J1713.7-3946 and alike can be resolved. It is anticipated that observations by \textit{FORCE} will reveal the spatial distribution of the accelerated CRs in such SNRs, providing invaluable information not yet available from current observations to confront hydrodynamic and spectral models and constrain the progenitor nature and particle acceleration physics.  

\subsection{Caveats}\label{subsec:MFA}

\begin{figure}[ht]
    \gridline{\fig{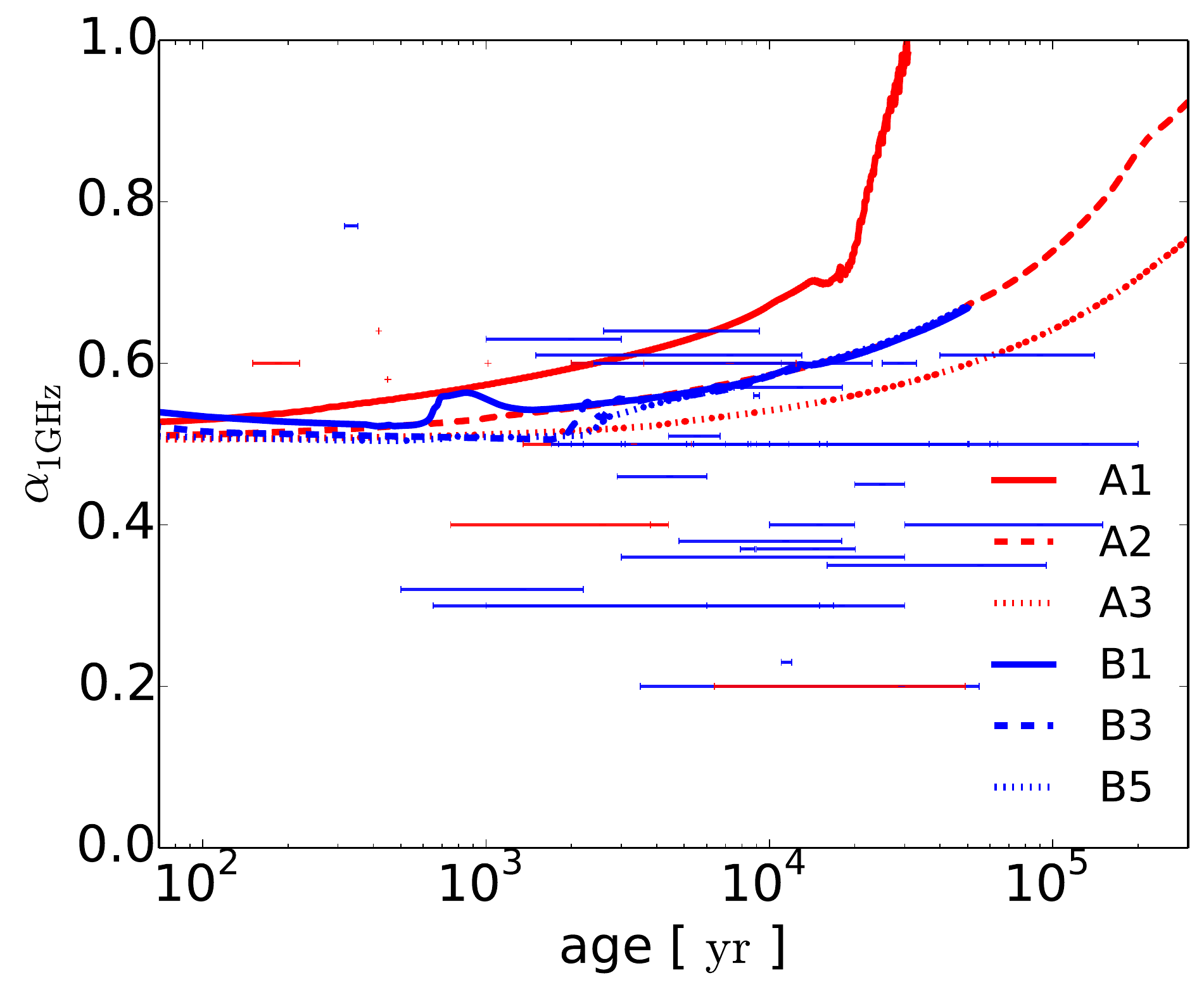}{0.4\textwidth}{(a) Radio spectral index at 1 GHz}}
    \gridline{\fig{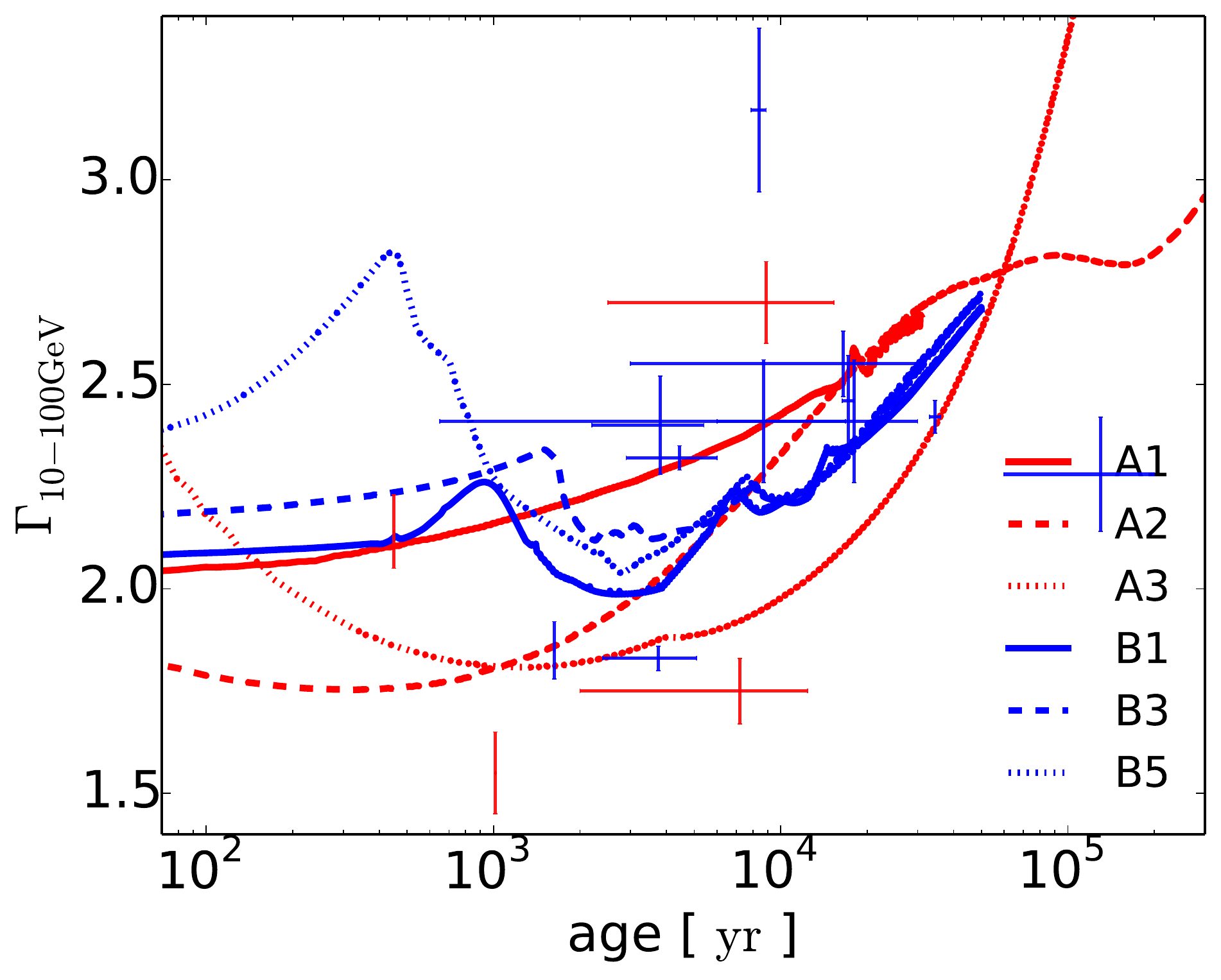}{0.4\textwidth}{(b) Gamma-ray index $\Gamma$ in 10--100 GeV}}
    \caption{Spectral index evolution at 1 GHz and 10--100 GeV. We denote $\alpha$ as in $S_\nu\propto\nu^{-\alpha}$ where $S_\nu$ is the flux density at frequency $\nu$, and $\Gamma$ is defined by $dN/dE\propto E^{-\Gamma}$. Indices from observations are overlaid as data points for reference, with the red and blue points for Ia and CC remnants respectively.}
        \label{fig:a-pl}
\end{figure}

In this section, we will elaborate on a few aspects which have not been fully considered or discussed within the scope of this work, as well as some possible future improvements on our models. 

As proposed in some scenarios, bright gamma-ray emission from evolved SNRs have been interpreted to be partially coming from the interaction between the escaped CRs and the surrounding dense clouds (at a distance from the SNRs), which appears to be successful in explaining the gamma-ray emission from RX J1713.7-3946 \citep{2018AA...612A...6H} and W28 \citep{2014ApJ...786..145H} for example. Focusing on the EM emission from the CRs confined inside the SNRs, this “illuminated clouds” emission component is currently beyond our scope and thus not included in this work. We do expect an increase of detected samples of gamma-ray emission associated with escaped CRs around SNRs from future observations by e.g., the \textit{Cherenkov Telescope Array}. We plan to include such a component as well as a more detailed discussion on the difference between the CR escape model \citep[e.g.,][]{2019MNRAS.487.3199C} and the Alfv\'enic drift model (for the emission from the confined CRs) in a follow-up paper.

A trend of spectral hardening has been observed in low-energy with SNR age \citep[][Figure 3(b)]{2019ApJ...874...50Z}, which is not well represented by our results (Figure~\ref{fig:a-pl}(a)). This can be stemming from the following reasons. As \citet{2019ApJ...874...50Z} suggested, the spectral hardening effect in older SNRs can be caused by several reasons, e.g., re-acceleration of the CR electron, Coulomb collision loss, and a secondary stochastic acceleration in the downstream. In fact, \citet{2015ApJ...806...71L} has used a framework similar to ours to investigate the radio spectral evolution in evolved SNRs interacting with molecular clouds, which seems partially responsible for the hardening trend in \citet{2019ApJ...874...50Z}. They found that a fast radiative cloud shock embedded inside a dense ($n\gtrsim 200\ \mathrm{cm^{-3}}$) cloud which re-accelerates Galactic CR electrons can indeed successfully explain the radio spectral hardening in old SNRs \citep[][see the left side of Figure 3(a)]{2015ApJ...806...71L}. Such emission including the gamma-rays are mainly contributed by a dense radiatively cooled shell right behind the FS, whereas the CRs trapped in the now tenuous interior of the SNR has already suffered from considerable adiabatic loss to become unimportant players for the overall spectrum at old ages. The difference of the 2015 work and this paper is that the parameter space surveyed in our models here (with a maximum $n=10\ \mathrm{cm^{-3}}$ and uniformly distributed ISM-like ambient medium) does not probe such a “crushed cloud” situation and hence cannot reproduce the spectral hardening result. Since the included physics are almost identical, an expansion of our parameter space in a follow-up work will be able to show such an effect. However, we note that older SNRs shown in \citet{2019ApJ...874...50Z} are not always interacting with dense clouds, and the crushed cloud scenario can only be a partial explanation to the spectral hardening effect witnessed in older SNRs in general. 

In the ``Alfv\'enic-drift" formalism we have adopted \citep[e.g.,][]{2001MNRAS.321..433B,CBAV2009a}, the magnetic waves are assumed to have a velocity the same as that of the local Alfv\'en waves. The effective compression ratio $S_\mathrm{sub}$ as written in Section~\ref{sec:methods} leads to a spectral softening with time (see also \citet{CBAV2009a}) as we can see in Figure~\ref{fig:a-pl}(b). We have not considered the effect suggested by \citet{2003AA...409..821V} that the magnetic waves in the downstream can be dominated by an outgoing component such that $S_\mathrm{sub}=(u_1-v_{A,1})/(u_2-v_{A,2})$ can be the case instead. As in \citet{CBAV2009a}, we have considered particle-wave interaction only in the upstream of the shock and ignored those in the downstream. In addition, following the approach of \citet{CBAV2009a,2010MNRAS.407.1773C}, we are using an approximated scheme for the magnetic-field amplification process in the shock precursor by extrapolating the quasi-linear regime with resonant scatterings only to a highly turbulent (Bohm) situation. A more self-consistent scheme is desirable for the treatment of particle-wave interactions, including the downstream regions. The behavior of magnetic turbulence and particle-wave interactions has been investigated through various paths, e.g., particles-in-cell and hybrid simulations, MHD simulations, Monte-Carlo approach and so on \citep{2013MNRAS.430.2873R,BEOV2014,2018MNRAS.473.3394V,2021ApJ...922....7I}, but the synergy with a long-term global hydrodynamic evolution with NLDSA applicable to the interpretation of broadband emission is still lacking. While we have restricted ourselves to the current more simplistic implementation of MFA in our models, we are on our way to incorporate a more sophisticated scheme to our framework and the results will be reported in a future work.  


\section{Summary} \label{sec:summary}

We have performed long-term simulations to study the evolution of non-thermal emission from SNRs in various kinds of environments. 
To realize these simulations, we adapt the \textit{CR-Hydro} code from \citetalias{2019ApJ...876...27Y} to our purposes by the implementation of several physical effects particularly relevant for SNRs in the radiative phase, such as a computationally efficient exact-integration scheme for radiative cooling, the re-acceleration of pre-existing CR populations, and a scheme for ion-neutral damping of magnetic waves. We studied two groups of models with a Type-Ia ejecta expanding into a purely uniform ambient medium (Group A) and with an ejecta from a RSG star surrounded by a CSM structure created by the pre-SN stellar wind (Group B), respectively. We analyzed the characteristics of the hydrodynamic evolution, multi-wavelength light curves and spectral evolution for each model and discussed on their dependence on the diversified ambient environment. Compared to \citetalias{2019ApJ...876...27Y}, we extend their calculation self-consistently to an age way past the onset of the radiative phase ($t = 3 \times t_\mathrm{tr}$) and follow the consequence on the non-thermal emission at the late time evolutionary stage of a SNR. The main results can be summarized as follows.

\begin{enumerate}
\item Results from models in Group A are found to be in agreement with \citetalias{2019ApJ...876...27Y} for the first 5,000 yrs, as well as for the first 1,000 yrs or so for Group B when the shock is propagating in a simple $\rho_0 \propto r^{-2}$ wind, confirming the robustness of our calculations. 

\item The non-thermal spectral evolution from the Sedov phase to the radiative phase can now be followed coherently in a common platform, improving on the ad hoc treatments adopted by previous studies which only focused on the local behavior of radiative shocks in a dense cloud. 

\item Characteristic spectral steepening is witnessed across the electromagnetic spectrum to various degrees for all models in the radiative phase due to a rapid decrease of the Alfv\'enic Mach number and hence the effective compression ratio of the shock, consistent with recent radio and gamma-ray observations of evolved SNRs. 

\item Depending on the mass-loss history and ejecta mass of the progenitor, the non-thermal spectrum of a CC SNR can ``lose memory'' from the past, i.e., after a few $10^3$ yrs the SED no longer retain any information of the CSM structure around the ejecta, and gradually converge to a homologous evolutionary track very similar to that without any density features created by the stellar wind. Exceptions are expected for stripped envelope SNRs with an enhanced mass-loss.

\item We investigated the age dependence of the importance of the re-acceleration of pre-existing Galactic CRs in terms of the long-term CR production history of a SNR in various environments. The fractional energy contribution of the re-accelerated CRs to the total CR population inside a SNR rises with age in general. The maximum fraction is reached in the radiative phase and is found to be in the ballpark of a few $10\%$ depending on the ambient environment. This is far from a complete domination in contrast to the conclusions of previous studies which claimed that re-accelerated CRs alone are sufficient to explain the non-thermal emission properties of evolved SNRs. The implication is that even in the radiative phase when the shock is no longer strong enough to sustain efficient particle acceleration of the thermal particles, there exists a non-negligible contribution to the emission from CRs accelerated in the past. As a result, it is crucial to follow the SNR evolution coherently from the explosion to current days in order to obtain an accurate estimate of the energy budget of the CRs and hence the interpretation of the observed non-thermal emission.

\item A spectral break in the radiative phase from ion-neutral damping as predicted by some previous studies cannot be confirmed in the overall SED of our models. While a momentum break indeed appears locally at the radiative shock in some models, the volume-integrated SED is found to be dominated by the CRs accelerated before the ion-neutral damping effect becomes important. A future study involving shock interaction with dense molecular clouds as well as a more realistic spatial structure of the environment may yield model spectra in which a clear spectral break can be observed.  

\item We also assessed the prospect of \textit{FORCE} on the study of non-thermal SNRs in the near future. In the 10--40~keV band, most of the emission in our models is dominated by the non-thermal bremsstrahlung and IC components. For models with a higher CSM/ISM density, we predict that \textit{FORCE} detection together with gamma-ray observations will be able to constrain the crucial electron-to-proton number ratio ($K_\mathrm{ep}$) at relativistic energies to help us understand the poorly understood electron injection and acceleration mechanism at strong collisionless shock. The superior angular resolution and large effective area of \textit{FORCE} will allow for space-resolved spectroscopy of extended non-thermal SNRs in the important hard X-ray band, which is essential for revealing any inhomogeneous distribution of CR protons and electrons inside the remnant as well as providing constraints on key parameters like the magnetic field strength. 

\end{enumerate}

This work has established a robust platform for simulating the long-term evolution of non-thermal emission from SNRs interacting with various types of CSM/ISM environments. We plan to implement stellar evolution models for different types of progenitor stars and their associated CSM structures in our next step \citep[e.g.,][]{Yasuda_2022}, so that we can provide a systematic survey on a rich diversity of SNR models for comparison with observation data from new missions such as \textit{FORCE}, \textit{CTA} and so on in the near future. 
Further improvements are underway as described in Section~\ref{subsec:MFA} on aspects such as the inclusion of the contribution from the escaped CRs and a revision on the particle-wave interaction scheme in the NLDSA framework.  
Another line of studies focusing on the thermal aspect of SNR emission is also underway in parallel \citep[e.g.,][and reference therein]{2018ApJ...865..151M,Jacovich_2021}. We plan to join effort with these thermal emission studies in the near future to construct a comprehensive model for multi-wavelength emission from Type Ia and CC SNRs.  

\begin{acknowledgements}
S.H.L. acknowledges support by JSPS grant No. JP19K03913 and the World Premier International Research Center Initiative (WPI), MEXT, Japan. H.Y. acknowledges support by JSPS Fellows grant No. JP20J10300.
\end{acknowledgements}


\bibliography{ms}{}
\bibliographystyle{aasjournal}

\clearpage
\begin{longrotatetable}
\begin{deluxetable*}{l|llclcccccl}
\centering
\tablecolumns{11}
\tablewidth{15cm}
\tablecaption{Observation data} 
\tablehead{
SNR & common name &type\tablenotemark{a} &age &distance &$r_\mathrm{sk}$ &$v_\mathrm{sk}$ &$F_\mathrm{1 GHz}$ &$F_\mathrm{1-100 GeV}$\tablenotemark{b} &$F_\mathrm{1-10 TeV}$\tablenotemark{c} &Ref\\
  &&& [yr] & [kpc] & [deg] & [''/yr] & [Jy] & [$10^{-9}\mathrm{cm^{-2}s^{-1}}$] &[$10^{-13}\mathrm{cm^{-2}s^{-1}}$]&   
}
\startdata
G1.9+0.3&&Ia& 150-220&8.5&0.014&0.35& 0.6&0.27&0.81&[1]\\
G4.5+6.8&Kepler, SN1604&Ia& 418&2.9-4.9&0.029&0.22$\pm$0.009& 19&0.32$\pm$0.09&0.69$\pm$0.17&[2]\\
G6.4-0.1&W28&CC& 33000-36000&1.6-2.2&0.6&0.0033$\pm$0.0003& 310&72.19$\pm$2.84&2.9$\pm$0.1&[3]\\
G7.7-3.7&&CC& 500-2200&3.2-6&0.15&-& 11&0.89&6.1&[4]\\
G15.9+0.2&&CC& 1000-3000&8.5&0.043$\pm$0.001&0.021$\pm$0.0007& 5.0&4.09$\pm$0.57&2.6&[5]\\
G18.1-0.1&&CC& 5100-9000&5.6-6.6&0.092$\pm$0.004&0.025$\pm$0.003& 4.6&3.5&9.46$\pm$0.65&[6]\\
G21.8-0.6&&CC& 8800-9200&5.4-5.8&0.145&0.049& 65&2.4&-&[7]\\
G23.3-0.3&W41&CC& 60000-200000&4.6-5&0.55&-& 70&8.11$\pm$0.86 &24.09$\pm$2.15&[8]\\
G33.6+0.1&Kes 79&CC& 4400-6700&3.5-7.1&0.083&0.015$\pm$0.001& 20&5.15$\pm$0.34&-&[9]\\
G34.7-0.4&W44&CC& 7900-8900&2.1-3.3&0.52$\pm$0.07&0.010$\pm$0.0002& 240&154.41$\pm$13.61&11.2&[10]\\
G41.1-0.3&3C397&Ia& 1350-5300&8-9&0.032$\pm$0.003&0.032$\pm$0.008& 25&7.7&-&[11]\\
G43.3-0.2&W 49B&CC& 2900-6000&10.9-11.7&0.033&0.021& 38&18.79$\pm$1.45 &1.46$\pm$0.21&[12]\\
G49.2-0.7&W51C&CC& 16400-18000&4.8-6&0.057&0.0034$\pm$0.0002& 160&36.08$\pm$2.80 &-&[13]\\
G53.6-2.2&&CC& 15000-50700&2.3-6.7&0.25$\pm$0.02&0.020$\pm$0.006& 8&64.37$\pm$1.49&5.3&[14]\\
G57.2+0.8&&CC& 16000-95000&5.9-7.3&0.092&-& 1.8&0.59&5.5&[15]\\
G65.1+0.6&&CC& 40000-140000&9-9.6&0.58$\pm$0.16&-& 5.5&2.8&19.4&[16]\\
G67.7+1.8&&CC& 1500-13000&7-17&0.08&0.018$\pm$0.005& 1.0&0.43&15.3&[17]\\
G73.9+0.9&&CC& 11000-12000&0.5-4&0.23&0.0047& 9&1.46$\pm$0.74&-&[18]\\
G74.0-8.5&Cygnus Loop, W78&CC& 10000-20000&0.576-1&1.6$\pm$0.2&0.077$\pm$0.047& 210&10.60$\pm$0.60 &-&[19]\\
G82.2+5.3&&CC& 14100-20900&1.3-3.2&1.2$\pm$0.6&0.069$\pm$0.002& 120&1.3&-&[20]\\
G84.2-0.8&&CC& 8400-11700&4.8-6.2&0.15$\pm$0.02&0.027$\pm$0.001& 11&0.34&-&[21]\\
G85.4+0.7&&CC& 3500-55000&2.5-5.2&0.3&0.11$\pm$0.02& -&2.2&-&[22]\\
G85.9-0.6&&Ia& 6400-49000&3.2-6.4&0.2&0.029$\pm$0.004& -&2.1&-&[23]\\
G89.0+4.7&&CC& 4800-18000&0.8-2.1&1.0&$<0.026$& 220&8.37$\pm$0.66&-&[24]\\
G111.7-2.1&Cas A&CC& 316-352&3.3-3.7&0.043$\pm$0.003&0.47& 2300&6.25$\pm$0.42 &5.8$\pm$1.2 &[25]\\
G116.5+1.1&&CC& 15000-50000&1.6&0.59&0.026& 10&1.8&-&[26]\\
G116.9+0.2&&CC& 7500-18100&1.6-3.5&0.30&0.0023$\pm$0.0004& 8&1.5&-&[27]\\
G120.1+1.4&Tycho, SN1572&Ia& 450&1.7-5&0.07&0.30$\pm$0.06& 50&0.91$\pm$0.16&1.1$\pm$0.4&[28]\\
G127.1+0.5&&CC& 20000-30000&1.15&0.44&-& 12&2.1&-&[29]\\
G132.7+1.3&&CC& 25000-33000&2-2.2&0.67&0.034$\pm$0.005& 45&6.3&-&[30]\\
G156.2+5.7&&CC& 7000-36600&1.7-3&0.9&0.035& 5&2.0&-&[31]\\
G160.9+2.6&&CC& 2600-9200&0.3-1.2&1.0&-& 110&2.8&-&[32]\\
G166.0+4.3&&CC& 9000-20100&1-4.5&0.38&0.015$\pm$0.010& 7&1.7&-&[33]\\
G182.4+4.3&&Ia& 3800-4400&$\geq$3&0.42&0.16& 0.5&0.82&-&[34]\\
G189.1+3.0&IC443, Jellyfish Nebula&CC& 3000-30000&0.7-2&0.38&0.012&165&83.09$\pm$4.18 &-&[35]\\
G205.5+0.5&&CC& 30000-150000&0.9-1.98&1.88&0.0059& 140&2.38$\pm$0.41&4.16$\pm$0.51&[36]\\
G260.4-3.4&Puppis A&CC& 2200-5400&1.3-2.2&0.6&0.12& 130&16.53$\pm$1.43 &-&[37]\\
G266.2-1.2&Vela Jr., RX J0852.0-4622&CC&2400-5100&0.5-1&0.86$\pm$0.17&0.42$\pm$0.10&50&11.53$\pm$0.76&274.62$\pm$14.04 &[38]\\ 
G272.2-3.2&&Ia& 3600-11000&2.5-5&0.13&0.032& 0.4&1.2&5.6&[39]\\
G290.1-0.8&&CC& 10000-20000&3.5-11&0.25&-& 42&0.46&5.8&[40]\\
G296.1-0.5&&CC& 2800-28000&2&0.26$\pm$0.05&0.022& 8&2.0&3.6&[41]\\
G296.5+10.0&&CC& 7000-10000&1.3-3.9&0.65$\pm$0.10&0.11$\pm$0.04& 48&0.94$\pm$0.16 &-&[42]\\
G296.8-0.3&&CC& 2000-11000&9&0.099&-& 9&1.4&5.7&[43]\\
G299.2-2.9&&Ia& 4500-11400&5&0.13&0.049& 0.5&0.76&71.8&[44]\\
G304.6+0.1&&CC& 2000-64000&$\geq$9.7&0.067&0.20$\pm$0.18& 14&3.6&-&[45]\\
G306.3-0.9&&Ia& 2500-15300&8&0.031&0.0074& 0.16&0.93$\pm$0.17&1.1&[46]\\
G308.4-1.4&&CC& 2400-7500&9.1-10.7&0.070&0.016$\pm$0.004& 0.4&0.58&3.5&[47]\\
G309.2-0.6&&CC& 700-4000&2-6&0.11$\pm$0.01&-& 7&0.96&3.8&[48]\\
G315.4-2.3&RCW 86&Ia& 2000-12400&2.3-3.2&0.35&0.12$\pm$0.06& 49&1.37$\pm$0.17&18.2$\pm$9.4&[49]\\
G327.4+0.4&Kes 27&CC& 2400-23000&4.3-6.5&0.17&0.028& 30&3.4&5.1&[50]\\
G327.6+14.6&SN1006&Ia& 1016&1.6-2.2&0.25&0.49& 19&0.088$\pm$0.041&3.7$\pm$0.8 &[51]\\
G330.0+15.0&&CC& 15000-52200&0.15-0.5&2.5$\pm$0.3&0.20$\pm$0.03& 350&0.48&-&[52]\\
G330.2+1.0&&CC& 1000-15000&$\geq$5&0.083&0.38& 5&0.64&8.2&[53]\\
G332.4+0.1&&CC& 3000-8600&7.5-11&0.13&0.14& 26&0.89&76.08$\pm$5.24&[54]\\
G337.2-0.7&&Ia& 750-4400&2-9.3&0.05&-& 1.5&1.1&-&[55]\\
G337.8-0.1&&CC& 1700-16000&12.3&0.063$\pm$0.013&0.0042$\pm$0.0002& 15&5.21$\pm$0.44&-&[56]\\
G344.7-0.1&&CC& 3000-6000&6.3-14&0.083&0.037& 2.5&4.5&40.88$\pm$3.31&[57]\\
G346.6-0.2&&Ia& 4200-16000&5.5-11&0.068$\pm$0.004&0.028& 8&1.7&-&[58]\\
G347.3-0.5&RX J1713.7-3946&CC& 1629&1&0.46$\pm$0.04&0.82$\pm$0.06& 30&8.20$\pm$0.64 &145.71$\pm$5.77&[59]\\
G348.5+0.1&&CC& 6000-30000&6.3-12.5&0.125&-& 72&15.61$\pm$1.75 &2.04$\pm$0.29&[60]\\
G348.7+0.3&&CC& 650-16800&9.8-13.2&0.085&0.020$\pm$0.002& 26&15.61$\pm$1.75&5.11$\pm$0.69&[61]\\
G349.7+0.2&&CC& 1800-3100&11.5-12&0.033&0.013& 20&4.00$\pm$0.76 &1.23$\pm$0.21&[62]\\
G355.6-0.0&&CC& 7300-20000&13&0.083$\pm$0.016&-& 3&3.3&2.7&[63]\\
\enddata
\tablecomments{For radio flux we use data from \citet{2019JApA...40...36G}, GeV flux data from \citet{2016ApJS..224....8A}, TeV flux and distance data from  \citet{2018AA...612A...3H} and references therein, other data from \textit{SNRcat} (\url{http://www.physics.umanitoba.ca/snr/SNRcat} \citep{2012AdSpR..49.1313F}), and the others: 
\textit{Fermi}:[2]\citet{2021ApJ...908...22X}[3,8,10,13,35,37,38,49,59,60,61]\citet{2017ApJ...843..139A}[5]\citet{2021arXiv210307824X}[9]\citet{2022ApJ...928...89H}[12]\citet{2018AA...612A...5H}[14]\citet{2017ApJ...842...22E}[18]\citet{2016MNRAS.455.1451Z}[28]\citet{2017ApJ...836...23A}[36]\citet{2017ApJ...846..169L}[42]\citet{2018ApJS..237...32A}[46]\citet{2017MNRAS.466.3434S}[51]\citet{2017ApJ...851..100C}[56]\citet{2018AA...619A.109S},
H.E.S.S.:[2]\citet{2022arXiv220105839H}[1,53]\citet{2014MNRAS.441..790H}[3]\citet{2020AA...633A.138S}[6]\citet{2020AA...644A.112H}[8,54,57,60,61]\citet{2018AA...612A...1H}[12]\citet{2018AA...612A...5H}[36]\citet{2014ApJ...780..168A}[38]\citet{2018AA...612A...7H}[59]\citet{2018AA...612A...6H}[62]\citet{2015AA...574A.100H}, 
$r_\mathrm{sk}$,$v_\mathrm{sk}$:[1]\citet{2011ApJ...737L..22C}[2]\citet{2008ApJ...689..231V}[3]\citet{2002AJ....124.2145V}[4]\citet{2018ApJ...865L...6Z}[5]\citet{2018MNRAS.479.3033S}[6]\citet{2014MNRAS.438.1813L}[7]\citet{2009ApJ...691..516Z}[8]\citet{2007ApJ...657L..25T}[9]\citet{2018ApJ...864..161K}[10]\citet{2013ApJ...777...14P}[13]\citet{2013ApJ...777...14P}[11]\citet{2016ApJ...817...74L}[12]\citet{2014ApJ...793...95Z,2007ApJ...654..938K}[14]\citet{1994AJ....108..207D,1983ApSS..89..279A}[15]\citet{2020ApJ...905...99Z}[16]\citet{2006AA...455.1053T}[17]\citet{2009AA...494.1005H}[18]\citet{2019JApA...40...36G,2016MNRAS.455.1451Z}[19]\citet{2018MNRAS.481.1786F}[20]\citet{1981RMxAA...5...93R}[21]\citet{2012ApJ...760...25L}[22]\citet{2008ApJ...674..936J}[23]\citet{2008ApJ...674..936J}[24]\citet{2007AA...461..991M}[25]\citet{2001ApJ...552L..39G,2009ApJ...697..535P}[26]\citet{2004ApJ...616..247Y,1981AA....99...17R}[27]\citet{2004ApJ...616..247Y,1980AA....84...26L}[28]\citet{2010ApJ...725..894H,2010ApJ...709.1387K}[29]\citet{1989AA...219..303J}[30]\citet{2006ApJ...647..350L}[31]\citet{1991AA...246L..28P,2009PASJ...61S.155K}[32]\citet{2007AA...461.1013L}[33]\citet{1970AuJPh..23..425M,1979AuJPh..32..113L}[34]\citet{1998AA...331..661K}[35]\citet{2019JApA...40...36G,2017MNRAS.472...51A}[36]\citet{1982AA...109..145G,1986ApJ...301..813O}[37]\citet{2017MNRAS.464.3029R}[38]\citet{2015ApJ...798...82A}[39]\citet{2016PASJ...68S...7K}[40]\citet{1996AA...315..243R}[41]\citet{2012MNRAS.419.1603G}[42]\citet{2019JApA...40...36G,1988ApJ...332..940R}[43]\citet{2012ApSS.337..573S}[44]\citet{2007ApJ...665.1173P,1996ApJ...465..840S}[45]\citet{2012MNRAS.423.1215G,2013ApJ...777..148G}[46]\citet{2019PASJ...71...61S}[47]\citet{2012AA...544A...7P}[48]\citet{2001ApJ...548..258R}[49]\citet{1996AA...315..243R,2013MNRAS.435..910H}[50]\citet{2008ApJ...676.1040C}[51]\citet{1988ApJ...332..940R}[52]\citet{1980AA....85..184T}[53]\citet{2004ApJ...604..693V}[54]\citet{}[55]\citet{2001ApJ...548..258R}[56]\citet{2008AA...488L..25C}[57]\citet{1996AAS..118..329W,2011AA...531A.138G}[58]\citet{2017ApJ...847..121A}[59]\citet{2016PASJ...68..108T}[60]\citet{2019JApA...40...36G}[61]\citet{2019MNRAS.487.5019B}[62]\citet{2014PASJ...66...68Y}[63]\citet{2013PASJ...65...99M}.
}
\tablenotetext{a}{For SN explosion type, `Ia' indicates type Ia explosion, and `CC' includes core-collapse explosion and explosion with inadequate information on the explosion type.}
\tablenotetext{b}{We use 1--100 GeV flux from \textit{Fermi}-LAT and for non-detected SNRs 99\% upper limits assuming spectral index $\Gamma=2.5$.}
\tablenotetext{c}{We use 1--10 TeV flux from H.E.S.S. and for non-detected SNRs 99\% upper limits assuming spectral index $\Gamma=2.3$.}
\tablenotetext{*}{For converting the observed photon number flux into the luminosity we use the same formula as Eq (12) in \citetalias{2019ApJ...876...27Y}.}
\label{tab:obs}
\end{deluxetable*}
\end{longrotatetable}



\end{document}